\documentclass[aps,pre,floatfix,twocolumn,nofootinbib]{revtex4}
\usepackage{graphicx} 
\usepackage[sort&compress]{natbib}
\usepackage{graphicx}
\usepackage{amsmath}\usepackage{amssymb}
\usepackage{subfigure}
\newcommand{\A}{a}
\newcommand{\ddelta}{\delta_{\cal D}}
\newcommand{\BEQ}{\begin{equation}}
\newcommand{\EEQ}{\end{equation}}
\newcommand{\BEA}{\begin{eqnarray}}
\newcommand{\EEA}{\end{eqnarray}}
\newcommand{\comment}[1]{}
\def\d {{ {\rm d} }}
\begin{document}
\draft

\title{Electromagnetic gauge-freedom and work}

\author{A.E. Allahverdyan and S.G. Babajanyan}

\address{Yerevan Physics Institute, Alikhanian Brothers Street
2, Yerevan 375036, Armenia }

\date{\today}

\begin{abstract} 
\comment{  We address the problem of defining the work done by a heavy body
  (source) on a lighter particle when the interaction between them is
  electromagnetic and relativistic. The straightforward solution of
  this problem is plagued by the fact that the time-dependent
  Hamiltonian of the light particle is not gauge-invariant. We
  reviewed previous proposals for resolving this problem and found
  them incomplete and/or unsatisfactory. We argue that a consistent
  definition of the work can be given via the gauge-invariant kinetic
  energy of the source that relates to the particle's Hamiltonian in
  the Lorenz gauge. The relation amounts to a formulation of the first
  law which is derived from relativistic dynamics and has definite
  validity conditions. It is based on a specific separation of the
  overall energy into those of the source, particle and electrodynamic
  field. This separation is deduced from a consistent energy-momentum
  tensor. Hence it holds relativistic covariance and causality.}

We argue that the definition of the thermodynamic work done on a
charged particle by a time-dependent electromagnetic field is an open
problem, because the particle's Hamiltonian is not
gauge-invariant. The solution of this problem demands accounting for
the source of the field. Hence we focus on the work done by a heavy
body (source) on a lighter particle when the interaction between them
is electromagnetic and relativistic. The work can be defined via the
gauge-invariant kinetic energy of the source. We uncover a formulation
of the first law (or the generalized work-energy theorem) which is
derived from relativistic dynamics, has definite validity conditions,
and relates the work to the particle's Hamiltonian in the Lorenz
gauge. Thereby the thermodynamic work also relates to the mechanic
work done by the Lorentz force acting on the source.  The formulation
of the first law is based on a specific separation of the overall
energy into those of the source, particle and electromagnetic
field. This separation is deduced from a consistent energy-momentum
tensor. Hence it holds relativistic covariance and causality.

\comment{ }

\end{abstract}








\maketitle

\section{Introduction}
\label{introduction}

Equilibrium statistical thermodynamics is based on notions of work, heat,
entropy and temperature \cite{balian,lindblad,mahler}. The primary
concept of non-equilibrium statistical mechanics is work, because its
definition is relatively straightforward
\cite{balian,lindblad,mahler}. This is witnessed by recent activity in
non-equilibrium (classical and quantum) physics that revolves around
the work and the laws of thermodynamics \cite{lindblad,mahler,lehrman,
mukamel,jarb,leff,campisi,skr,armen,gallego,aspects,plastina,3law,
  rubi,peliti,silbey,rubirubi,jar}.

We recall the definition of the thermodynamic work and its features.
Consider a non-relativistic particle with coordinate $x$, canonic
momentum $\pi$ and Hamiltonian ${\cal H}(x,\pi;f(t))$, where $f(t)$ is
an external field. The thermodynamic work done on the particle by the
field's source in the time-interval $[t_1,t_2]$ is
\cite{balian,lindblad,mahler}:
\begin{eqnarray}
  \label{eq:00}
W=  {\cal H}(x(t_2), \pi(t_2);f(t_2))-{\cal H}(x(t_1), \pi(t_1);f(t_1)).
\end{eqnarray}
$W$ equals the energy increase of the particle. No work is done if $f$
is time-independent. Definition (\ref{eq:00}) generalizes to
statistical situations, where the description goes via probability
densities or via density matrices \cite{balian,lindblad,mahler}. It
appears in the laws of thermodynamics.

There is an alternative definition of the thermodynamic work
\cite{balian,lindblad,mahler} 
\begin{eqnarray}
  \label{eq:43}
W= \int_{t_1}^{t_2}\d t\,
\frac{\d f}{\d t}\,\partial_f {\cal H} (x(t), \pi(t); f(t)).  
\end{eqnarray}
It applies to the open-system situation, e.g. particles interacting
with baths \cite{balian,lindblad,mahler}. Eq.~(\ref{eq:43}) leads to
(\ref{eq:00}) due to the Hamilton equations of motion. 

If $f$ is a coordinate of the source, the full time-independent
Hamiltonian of the system and source reads\footnote{Even if $f$ is
  not a coordinate, (\ref{eq:00}, \ref{eq:43}) stay consistent as far
  as the system and the work-source form a closed system. }
\begin{eqnarray}
  \label{eq:740}
{\cal H}(x,\pi;f)+
  {\cal H}'(f,\pi_f),
\end{eqnarray}
where $\pi_f$ is the momentum of the source, and ${\cal H}'(f,\pi_f)$
is its Hamiltonian. The time-dependent Hamiltonian ${\cal
  H}(x,\pi;f(t))$ of the system in (\ref{eq:00}, \ref{eq:43}) results
from (\ref{eq:740}), if the reaction of the system to the source is
neglected, e.g. because the source is heavy.

Two important features of the thermodynamic work (\ref{eq:00}) are
displayed via (\ref{eq:43}, \ref{eq:740}). First, since the total
energy (\ref{eq:740}) is conserved, the thermodynamic work
(\ref{eq:00}) relates to the energy change ${\cal H}'(f,\pi_f)$ of the
source \cite{balian,lindblad,mahler}. Second, (\ref{eq:43},
\ref{eq:740}) show that $-\partial_f {\cal H} (x(t), \pi(t); f(t))$ is
the potential force acting (from the particle) to the source. Then
(\ref{eq:43}) relates to the mechanic concept of work: force times the
displacement $\d t\, \frac{\d f}{\d t}$ \cite{leff}.

Let now the external field be electromagnetic (EMF). The relativistic
Hamiltonian of a particle reads \cite{landau,kos}
\begin{eqnarray}
  \label{eq:01}
H=\sqrt{c^2\,\vec{p}\,^2 
+m^2c^4   }+e\phi(\vec{x},t),\\
  \label{eq:02}
\vec{p}=m\vec{v}/\sqrt{1-v^2/c^2}, \qquad 
\vec{\pi}=\vec{p}+{e}\,\vec{A}(\vec{x},t)/c,
\end{eqnarray}
where $A^i=(\phi,\vec{A})$ is the 4-potential of EMF, $\vec{p}$
($\vec{\pi}$) is the kinetic (canonic) momentum, $\vec{x}$, $m$ and
$e$ are the coordinate, mass and charge, respectively.  For
a non-relativistic particle, $\sqrt{c^2\vec{p}\,^2 +m^2c^4}$ 
is replaced by $\frac{\vec{p}\,^2 }{2m}$.

The thermodynamic work cannot be read directly from (\ref{eq:00},
\ref{eq:43}, \ref{eq:01}), because neither $H$ nor its time-difference
stay invariant under a gauge-transformation defined via a function
$\chi(\vec{x},t$) \cite{yang,kobe,zambrano,lorce_review,grif}:
\begin{eqnarray}
  \label{eq:27}
A^i\to A^i+\partial^i \chi:  &&  \phi(\vec{x},t)\to \phi(\vec{x},t)
+{\partial \chi(\vec{x},t)}/{\partial(c t)}, ~~\\
&&  \vec{A}(\vec{x},t)\to \vec{A}(\vec{x},t)-
{\partial \chi(\vec{x},t)}/{\partial \vec{x}}.
\end{eqnarray}
The kinetic momentum $\vec{p}$ in (\ref{eq:01}) is
gauge-invariant. But $\phi$ is not \cite{yang,kobe} \footnote{ This
  differs from a formally similar gauge-freedom issue that appears
  even for the non-relativistic situation (\ref{eq:00})
  \cite{rubi,peliti,silbey,rubirubi}. This issue is resolved easily by
  choosing $f(t)$ as e.g. the coordinate of a physical source of work
  \cite{rubi,peliti,silbey,rubirubi,jar}; see (\ref{eq:740}) and 
Appendix \ref{uno}. }.
The same problem exists for a quantum particle interacting with EMF
\cite{yang,kobe}.

One response to the EMF gauge-freedom problem is that the gauge in
(\ref{eq:01}) is to be selected as (temporal gauge)
\cite{landau,kos,kobe}: 
\begin{eqnarray}
  \label{dalal}
\phi=0.
\end{eqnarray}
This definition is indirectly supported by the standard EMF
energy-momentum tensor, which suggests that particles do not have
potential energy \cite{landau,kos}; see Appendix
\ref{standard_standard}. Eq.~(\ref{eq:01}) under $\phi=0$ implies that
the sought thermodynamic work would amount to the particle's kinetic
energy change, i.e. to the mechanic work done on the particle by the
Lorentz force \cite{landau}. This cannot be the correct definition of
the thermodynamic work. First, because it implies that
time-independent fields do thermodynamic work \cite{reiss_1,reiss_2}
\footnote{Consider a constant electric field
  $\vec{E}=\partial_{\vec{x}}\phi(\vec{x})$. The temporal gauge is
  achieved by taking $\chi=-ct\phi(\vec{x})$ in (\ref{eq:27}), which
  brings in a time-dependent $\vec{A}(\vec{x},t)$ and shows from
  (\ref{eq:00}, \ref{eq:01}) that the work done is not zero and equals
  to the change of the kinetic energy. The latter is non-zero, since
  {\it only} $\sqrt{c^2\,\vec{p}\,^2 +m^2c^4 }+e\phi(\vec{x})$ is
  conserved in time \cite{landau}.}.  Second, because this work
(\ref{eq:43}) relates to the force acting on the source, and not to
the force acting on the particle.

Another possibility for resolving the gauge-freedom is to employ in
(\ref{eq:01}) the Coulomb gauge \cite{sipe,wang_pra}:
\begin{eqnarray}
  \label{eq:60}
  {\rm div}\vec{A}(\vec{x},t)=0.
\end{eqnarray}
But the scalar potential $\phi_{\rm C}(\vec{x},t)$ in the gauge
(\ref{eq:60}) propagates with infinite speed
\cite{stewart,heras_coulomb,wundt} \footnote{The Maxwell equation
  ${\rm div}\vec{E}=4\pi\rho$ reads \cite{landau}: $\Delta \phi
  (\vec{x},t)+\frac{1}{c}\partial_{t} {\rm
    div}\vec{A}(\vec{x},t)=-4\pi\rho(\vec{x},t)$. Upon using
  (\ref{eq:60}) we get for time-dependent $\phi_{\rm C}(\vec{x},t)$ in
  the Coulomb gauge the ``static'' equation $ \Delta \phi_{\rm C}
  (\vec{x},t)=-4\pi\rho(\vec{x},t)$ implying that $\phi_{\rm
    C}(\vec{x},t)$ responds instantly to changes in the charge density
  $\rho(\vec{x},t)$ \cite{stewart,heras_coulomb,wundt}}. While this is
consistent with electric $\vec{E}(\vec{x},t)$ and magnetic
$\vec{B}(\vec{x},t)$ fields propagating with speed $c$
\cite{stewart,heras_coulomb,wundt}, it also means that $\phi_{\rm
  C}(\vec{x},t)$|and the Hamiltonian (\ref{eq:01}) defined via
it|cannot be given a direct physical meaning.

For both proposals it is unclear how the work defined via
(\ref{eq:01}) relates to the energy of the source (heavy body).

Noting the difficulties with the temporal and Coulomb gauge, it is
natural to look at the Lorenz gauge,
\begin{eqnarray}
  \label{eq:13}
  \partial_iA^i\equiv\frac{1}{c}\frac{\partial \phi}{\partial t}+{\rm
    div}\vec{A}=0,
\end{eqnarray}
which is relavistic invariant and causal, i.e. $\phi$ and $\vec{A}$
defined from (\ref{eq:13}) propogate with speed $c$
\cite{heras_lorenz,dmitro}. Given (\ref{eq:13}) and standard boundary
conditions of decaying at spatial infinity, $A^i$ is uniquely
expressed via gauge-invariant electromagnetic field
$(\vec{E},\vec{B})$, and hence it is observable
\cite{heras_lorenz}. If the photon is found to have a small (but
finite) mass, (\ref{eq:13}) will hold automatically
\cite{massphoton}. These features are suggestive, but they do not
suffice for defining the thermodynamic work.

Our results validate the usage of (\ref{eq:01}, \ref{eq:13}), and also
show that main features of the thermodynamic work generalize to
the relativistic, electromagnetic situation:

-- The definition of the (thermodynamic) work based on (\ref{eq:01},
\ref{eq:13}) results from a separation of overall (source +
particle(s) + EMF) energy into specific components. This separation is
not arbitrary, but emerges from a relativistically covariant
energy-momentum tensor $\mathbb{T}^{ik}$ for the overall system; see
section \ref{tensor}. $\mathbb{T}^{ik}$ necessarily differs from the
standard energy-momentum tensor (e.g. because $\mathbb{T}^{ik}$ has to
account for a potential energy), but it leads to the same values of
the overall energy. It consistently relates to an angular momentum
(tensor). Certain aspects of $\mathbb{T}^{ik}$ are known from
\cite{puthoff,bogo}, but in its entirety it is proposed for the first
time \footnote{Energy-momentum tensors are not uniquely defined, and
  different situations may require different definitions. An example
  of this is the dielectric media (not considered here), where
  different experiments demand different forms of this tensor
  \cite{grif}.}.

-- The approach leads to a formulation of the first law for a
relativistic thermally isolated situation, which we demonstrate for
point charges with retarded electromagnetic interactions. According to
this formulation, the thermodynamic work can be defined through the
{\it gauge-invariant} kinetic energy of the source, but it is also
equal to the change of (\ref{eq:01}) in the Lorenz gauge. As compared
to the non-relativistic first law|which is an automatic consequence of
energy conservation \cite{lindblad,balian,mahler,lehrman,leff}
[cf. (\ref{eq:740})]|the formulation is necessarily approximate, since
some energy is stored in the (near) EMF, even if the radiated energy
is negligible. Once the thermodynamic work amounts to the kinetic
energy of the source, it directly relates it to the mechanic work done
by the Lorentz force acting on the source. Thus the two important
features (described after (\ref{eq:740})) generailze to the
relativistic, electromagnetic situation. 



Several recent studies looked at the work done by EMF in the context
of fluctuation theorems
\cite{deffner,uribe,saha,pradhan,yet_another}. But the problem of the
EMF gauge-freedom was not addressed, partially due to implicitly
assumed magnetostatic limit, where the gauge (\ref{eq:60}) is employed
by default, and where (\ref{eq:13}) and (\ref{eq:60}) are
approximately equal \cite{lar,rouss_epl}. 

This paper is organized as follows. Section \ref{tensor} derives from
equations of motion a new expression for the energy-momentum tensor of
EMF. Appendix \ref{standard_standard} compares with the standard
approach to this tensor. Section \ref{non_self_int} recalls the
relativistic dynamics of two point charges. This framework serves for
formulating the first law in section \ref{wowo}. Sections
\ref{selfself} shows that the formulation extends to certain
situations, where the radiation reaction is essential. The
last section summarizes and outlines open problems.

We use Gaussian units and metric $g_{ik}={\rm diag}
[1,-1,-1,-1]$. Vectors are denoted as $y^i=(y^0,\vec{y})$ and
$\vec{y}=(y^\alpha)$. We denote $x^i=(ct,\vec{x})$ and
$\partial_i\equiv\partial/\partial x^i$ for the 4-coordinate and
4-gradient, respectively.

\section{Equations of motion and energy-momentum tensor  }
\label{tensor}

Consider electromagnetic field (EMF) coupled with a charged continuous
matter with mass denisty $\mu(\vec{x},t)$, charge density
$\rho(\vec{x},t)$ and 4-velocity
\begin{eqnarray}
  \label{eq:33}
  u^i(\vec{x},t)=\frac{(1\,,\, {\vec{v}(\vec{x},t)}/{c}\,)}
{\sqrt{1-v^2(\vec{x},t)/c^2}}, \qquad
u_iu^i=1.
\end{eqnarray}
The comoving frame mass density and charge density
read, respectively (omitting $(\vec{x},t)$):
\begin{eqnarray}
  \label{meke}
  \mu_0=\mu\sqrt{1-\vec{v}^2/c^2}, \qquad
  \rho_0=\rho\sqrt{1-\vec{v}^2/c^2}.
\end{eqnarray}
Dealing with a continuous matter allows us to
postpone the treatment of infinities related to point particles.

The mass and charge conservation read, respectively
\begin{eqnarray}
  \label{eq:46}
  \partial_k (\mu_0u^k)=  \partial_k J^k =0,
\qquad J^k\equiv c\rho_0u^k.
\end{eqnarray}
where $J^k$ is the charge current. Equations of motion for
matter+EMF in the gauge (\ref{eq:13}) read \cite{landau,kos}
\footnote{Eqs.~(\ref{eq:22}, \ref{eq:31}) are normally obtained via
  (\ref{eq:13}). But one can employ (\ref{eq:22}) and (\ref{eq:46})
  for deriving the Lorenz gauge (\ref{eq:13})
  \cite{oost,fock,gritsunov}. }:
\begin{eqnarray}
  \label{eq:22}
&&  \partial_k \partial^k A^i=\frac{4\pi}{c}J^i, \\
  \label{eq:31}
&& \mu_0 c^2 \frac{\d u^i}{\d
  s}\equiv \mu_0 c^2 u^l\partial_l u^i =\frac{1}{c} F^{ik}J_k,
\end{eqnarray}
where $F^{ik}=\partial^i A^k- \partial^k A^i$, and $s/c$ is the proper
time.

The energy-momentum of matter reads \cite{landau}
 \begin{eqnarray}
  \label{eq:19}
  \tau_{i}^{\,\,k}=c^2\mu_0\,u_i\, u^k ,
\end{eqnarray}
where the pressure has been neglected.
Eq.~(\ref{eq:31}) implies
\begin{eqnarray}
\partial_k  \tau_{i}^{\,\,k} = \frac{1}{c}J^kF_{ik}.
  \label{eq:222}
\end{eqnarray}

We now deduce a conserved energy-momentum tensor $\mathbb{T}^{ik}$
from (\ref{eq:46}--\ref{eq:222}). Guided by the analogy with a free
scalar, massless field $\varphi$ whose energy-momentum tensor is
$\propto \partial^i\varphi\partial^k\varphi
-\frac{1}{2}g^{ik}\partial_l\varphi
\partial^l\varphi$ \cite{bogo}, we suggest 
\begin{eqnarray}
  \label{eq:20}
 && \mathbb{T}^{ik}= {T}^{ik}+\tau^{ik}+ \frac{1}{c}A^iJ^k,\\
 &&  T^{ik}
=-\frac{1}{4\pi}\, \partial^iA_l\, \partial^k A^l +\frac{1}{8\pi}\,
g^{ik} \,
\partial_nA_m\,\partial^n A^m,
  \label{eq:166a}
\end{eqnarray}
where (\ref{eq:166a}) is the energy-momentum tensor of the free EMF,
$\tau^{ik}$ is given by (\ref{eq:19}), and $\frac{1}{c}A^iJ^k$ in
(\ref{eq:20}) is due to the
interaction. Eqs.~(\ref{eq:46}--\ref{eq:222}) lead to 4 conservation
laws
\begin{eqnarray}
  \label{eq:21}
\partial_k  \mathbb{T}^{ik}=\partial_0  \mathbb{T}^{i0}
+\partial_\alpha  \mathbb{T}^{i\alpha}
=0.
\end{eqnarray}
Eqs.~(\ref{eq:21}, \ref{eq:20}) imply that 
$\mathbb{T}^{00}$ is the energy density
\begin{eqnarray}
  \label{eq:32}
  \mathbb{T}^{00}&=&-\frac{1}{4\pi}\partial^0A_l\partial^0A^l
+\frac{1}{8\pi}\partial_nA_m\partial^nA^m\\
&+&\frac{c^2\mu}{\sqrt{1-v^2/c^2}}+\rho\phi.
 \label{eq:322}
\end{eqnarray}
Eq.~(\ref{eq:32}) is the energy density of EMF, while (\ref{eq:322})
amounts to the energy of the matter that consists of kinetic and the
interaction term. The latter will be shown to be the particle's
potential energy in section \ref{wowo}. The possibility of this
interpretation is confirmed by the form of energy current:
\begin{eqnarray}
  \label{eq:34}
  c\mathbb{T}^{0\alpha}&=&-\frac{c}{4\pi}
\partial^0A_l\partial^\alpha A^l\\
&+& (\frac{c^2\mu}{\sqrt{1-v^2/c^2}} +\rho\phi)v^\alpha,
\label{eq:344}
\end{eqnarray}
where (\ref{eq:34}) is the energy current of EMF.

Eq.~(\ref{eq:166a}) for the free EMF was previously derived from the
Fermi's Lagrangian \cite{fermi,bogo,oost}; see Appendix
\ref{fermi_fermi}. Eqs.~(\ref{eq:32}--\ref{eq:344}) were discussed in
\cite{puthoff} as alternatives to standard expressions (i.e. the
Poynting vector), but were not derived from a consistent
energy-momentum tensor.  Various proposals for the energy flow of
(free) EMF are given in \cite{slepian}. Their general drawback is that
they do not start from a consistent energy-momentum tensor.

All above expressions|including $\mathbb{T}^{ik}$|that
contain $A^i$ (in the Lorenz gauge (\ref{eq:13})) are gauge-invariant,
because $A^i$ can be expressed via $F^{ik}=\partial^i A^k- \partial^k
A^i$ \cite{heras_lorenz}:
\begin{eqnarray}
  \label{eq:51}
&& A^i(x)=\partial_k\int\d^4 x'~ G(x-x')F^{ki}(x'),\\
  \label{eq:511}
&& \partial_k\partial^k G(x-x')=\delta_{\cal D}(x-x'),  \\
&& G(x-x')=\frac{1}{2\pi}\theta(x^0-x'^0)
\delta_{\cal D}((x^i-x'^i)(x_i-x'_i)),~~~~
\end{eqnarray}
where $G(x,x')$ is the retarded Green's function, $x\equiv
(ct,\vec{x})$, $x'\equiv (ct',\vec{x}')$,
$\theta(x^0-x'^0)=\theta(ct-ct')$ is the step function, and
$\delta_{\cal D}(x)$ is the Dirac's delta-function. Eq.~(\ref{eq:51})
relates to the retarded solution of (\ref{eq:22}). Its derivation from
(\ref{eq:511}) is straightforward, e.g. by employing (\ref{eq:13}) in
$F^{ik}$ and integrating by parts.

The main reason for introducing $\mathbb{T}^{ik}$ is to verify that
the potential energy $\rho\phi$ can emerge from a consistent
energy-momentum tensor; the standard energy-momentum tensor of EMF
does not allow such an interpretation; see Appendix
\ref{standard_standard}. Without a potential energy we cannot define
the thermodynamic work via (\ref{eq:00}, \ref{eq:43}); cf. the
discussion around (\ref{dalal}).

Both (\ref{eq:20}) and the standard tensor lead to the same
expressions for energy (and momentum) of the matter+EMF [see Appendix
\ref{radiation}]:
\begin{eqnarray}
  \label{eq:50}
  \int \d^3 x\,
  \left[ \frac{\vec{E}^2+\vec{B}^2  }{8\pi} +\tau^{00}\right]
  =\int \d^3 x\, \mathbb{T}^{00},
\end{eqnarray}
where the integration over the full 3-space is taken (assuming that
all fields nullify at infinity), $\frac{\vec{E}^2+\vec{B}^2 }{8\pi}$
is the Larmor's electromagnetic enegy density expressed via the
electromagnetic field $(\vec{E}, \vec{B})$ (it follows the standard
energy-momentum tensor \cite{landau}), and
$\tau^{00}=\frac{c^2\mu}{\sqrt{1-\vec{v}^2 }}$ is the kinetic energy
density for the matter; see (\ref{eq:19}, \ref{meke}). The same
$\tau^{00}$ enters also $\mathbb{T}^{00}$; cf. (\ref{eq:32},
\ref{eq:322}). Thus whenever only the total energy matters,
$\mathbb{T}^{00}$ agrees with the standard predictions
(\ref{eq:50}). However, generally the density
$\frac{\vec{E}^2+\vec{B}^2 }{8\pi}$ is not equal to $T^{00}$ given by
(\ref{eq:166a}). Differences and similarities between (\ref{eq:20})
and the standard energy-momentum tensor of EMF are discussed in
Appendices \ref{standard_standard} and \ref{radiation}; e.g.  for
spherical waves (\ref{eq:166a}) produces the same expression as the
standard tensor.

\comment{
$\frac{1}{c}\mathbb{T}^{\alpha 0}$ is the momentum density
\begin{eqnarray}
  \label{eq:35}
  \frac{1}{c}\mathbb{T}^{\alpha 0} &=&-\frac{1}{4\pi c}
\partial^0A_l\partial^\alpha A^l\\
&+& \frac{\mu v^\alpha}{\sqrt{1-v^2/c^2}} +\frac{\rho}{c} A^\alpha. 
  \label{eq:355}
\end{eqnarray}}

Note that the free EMF tensor (\ref{eq:166a}) is symmetric, ${T}^{ik}
={T}^{ki}$, as it should, because this ensures the known relation
between the energy current $cT^{0\alpha}$ and the momentum density
$T^{\alpha 0}=T^{0\alpha}$; cf. (\ref{eq:34}). But the full tensor
(\ref{eq:20}) is not symmetric, $\mathbb{T}^{ik}\not =\mathbb{T}^{ki}$,
due to the EMF-matter coupling. Appendix \ref{angu} discusses the
meaning of this asymmetry and relates $\mathbb{T}^{ik}$ to the angular
momentum and spin tensor. These relations are necessary to establish,
because the angular momentum is employed for explaining the energy of
EMF \cite{feynman,slepian,pugh}.

\section{Two point-particles without self-interactions}
\label{non_self_int}

For two point particles ${\rm P}$ and ${\rm P}'$ we take in
(\ref{meke}--\ref{eq:31}):
\begin{eqnarray}
  \label{eq:36}
&&  \mu(\vec{r},t)=
 m \ddelta\left(\vec{r}-\vec{x}(t)\right)
+ m' \ddelta\left(\vec{r}-\vec{x}'(t)\right), \\
&&  \rho(\vec{r},t)=
e \ddelta\left(\vec{r}-\vec{x}(t)\right)
+ e' \ddelta\left(\vec{r}-\vec{x}'(t)\right),
  \label{eq:366}
\end{eqnarray}
where $\vec{x}(t)$, $e$ and $m$ are the trajectory, charge and mass of
${\rm P}$ (resp. for ${\rm P}'$), and where $\ddelta$ is the
delta-function. It is known that for point particles equations of
motion (\ref{eq:31}) and energy-momentum tensor are not well-defined,
since they contain diverging terms \cite{landau,kos}.  One needs to
renormalize the masses by infinitely large counter-terms
\cite{kos}. The next-order (finite) terms refer to the self-force that
includes the back-reaction of the emitted radiation \cite{landau,kos}.

For clarity, we first focus on the point-particle case, where the
self-force is neglected, but the situation is still relativistic,
i.e. retardation effects are essential
\cite{synge,driver,driver_existence,
  stabler,cornish,leiter,beil,franklin,kirpich}. In this situation
particles influence each other via the Lorentz forces generated by the
Lienard-Wichert potentials; see Appendix \ref{deri}.

A sufficient condition for neglecting the self-force is that the
characteristic lengths are larger than the ``classical radius''
\cite{landau,synge,stabler,cornish,leiter,beil}
\begin{eqnarray}
  \label{eq:8}
{\rm
  max}[\,\frac{e^2}{m c^2 \,\sqrt{1-v^2/c^2}}, ~
\frac{e'^2}{m' c^2 \,\sqrt{1-v'^2/c^2}} ].  
\end{eqnarray}
Appendix \ref{deri} recalls how to get equations of motion for point
particles from (\ref{eq:31}) by selecting in (\ref{eq:22}) retarded
solutions, and relates them to the tensor (\ref{eq:20}).

We focus on the 1D situation, where the particles ${\rm P}$ and ${\rm
  P}'$ move on a line, since their initial velocities were collinear. We
checked that physical results obtained in sections
\ref{wowo}--\ref{selfself} hold as well for the full 3D situation, but
the 1D case is chosen for its relative simplicity. For all times $t_1$
and $t_2$, we set for the coordinates of ${\rm P}$ and ${\rm P}'$
(respectively)
\begin{eqnarray}
  \label{eq:0}
  x(t_1)\leq x'(t_2).
\end{eqnarray}
We denote for the delays $\delta(t)$ and $\delta'(t)$ that emerge due
to retarded interactions:
\begin{eqnarray}
  \label{eq:4}
  c\delta(t) \equiv x'-x(t-\delta(t)), \\
  c\delta'(t) \equiv x'(t-\delta'(t))-x,
  \label{eq:04}
\end{eqnarray}
and introduce dimensionless velocities: $\omega\equiv\dot{x}/c$,
$\omega'\equiv\dot{x}'/c$. The equations of motion read [see Appendix
\ref{deri}]
\begin{gather}
  \label{eq:5}
  \dot \omega (t)  = [1-\omega^2(t)]^{3/2}
\left(\frac{-ee'}{mc^3}\right)\frac{1}{\delta'^2(t)}\, 
\frac{1-\omega'(t-\delta'(t))}{1+\omega'(t-\delta'(t))},\\
  \label{eq:55}
  \dot\omega' (t)  = [1-\omega'^2(t)]^{3/2}
\left(\frac{ee'}{m'c^3}\right)\frac{1}{\delta^2(t)}\, 
\frac{1+\omega(t-\delta(t))}{1-\omega(t-\delta(t))}, \\
  \label{eq:9}
    \dot\delta' (t)
    =\frac{\omega'(t-\delta'(t))-\omega(t)}{1+\omega'(t-\delta'(t))}, \\
  \dot\delta (t)
    =\frac{\omega'(t)-\omega(t-\delta(t))}{1-\omega(t-\delta(t))}.
\label{eq:99}
\end{gather}
The factor $\delta'^{-2}(t)$ in (\ref{eq:55}) is the retarded Coulomb
interaction; cf. (\ref{eq:4}). Eqs.~(\ref{eq:5}--\ref{eq:99}) were
considered in \cite{synge,baylis,kasher,franklin}, and from a
mathematical viewpoint in
\cite{driver_book,driver,travis,hsing,zhdanov}. But the energy
exchange was not studied.

Eqs.~(\ref{eq:5}--\ref{eq:99}) are delay-differential equations due to
the retardation of the inter-particle interaction. Their initial
conditions are not trivial
\cite{kirpich,travis,hsing,zhdanov,aichelburg,book_numerics}. We
focus on the simplest scenario, where the two-particle system is
prepared via strong external fields for $t<0$
\cite{driver_book,book_numerics}. These fields do not enter into
(\ref{eq:5}--\ref{eq:99}) and they are suddenly switched off at the
initial time $t=0$. They define (prescribed) trajectories of the
particles for $t<0$. For simplicity we shall take them as
\begin{eqnarray}
  \label{eq:77}
x(t)= \omega_0\,c\,t, ~~
x'(t)= \omega_0' \,c\,t+l_0,~~ l_0>0, ~~ t\leq 0,
\end{eqnarray}
where $\omega_0$, $\omega'_0$ and $l_0$ are constants.  
Eqs.~(\ref{eq:4}, \ref{eq:04}, \ref{eq:77}) imply
\begin{eqnarray}
  \label{eq:45}
  \delta(0)=\frac{l_0/c}{1-\omega_0}, \qquad 
  \delta'(0)=\frac{l_0/c}{1+\omega'_0}.
\end{eqnarray}
Conditions (\ref{eq:77}, \ref{eq:45}) do determine uniquely the
solution of (\ref{eq:5}--\ref{eq:99}) for $t>0$
\cite{driver,driver_existence,driver_book,book_numerics}. An iterative
method of solving (\ref{eq:5}--\ref{eq:99}) is described in Appendix
\ref{nunu}. Section \ref{selfself} studies a different type of initial
conditions. 


\section{Work and the first law}
\label{wowo}

\subsection{Formulation of the first law}
\label{wowo_def}

The standard (non-relativistic) work has two aspects: the kinetic
energy change of the work-source (a heavy body whose motion is only
weakly perturbed by the interaction) \footnote{If the work-source in
  subject to an external potential, the latter will add to the kinetic
  energy.}, and the energy change of the lighter particle; cf. the
discussion around (\ref{eq:740}). The equality between them is the
message of the first law (in the thermally isolated situation, when no
heat is involved) \cite{balian,lindblad,mahler}. The task of
identifying two aspects of work will be carried out for the
relativistic dynamics (\ref{eq:5}--\ref{eq:99}).  To make ${\rm P'}$ a
source of EMF whose motion is perturbed weakly, we assume that it is
much heavier than ${\rm P}$:
\begin{eqnarray}
  \label{eq:57}
m'\gg m.
\end{eqnarray}
In practice, $m'/m\simeq 5-10$ already suffices; see below. 

The energy $E$ of ${\rm P}$ is defined via (\ref{eq:01}) in the Lorenz
gauge, or with help of (\ref{eq:322}):
\begin{eqnarray}
\label{eq:10}
E(t)&=&\frac{mc^2}{\sqrt{1-\omega^2(t)}}+e\phi'(x(t),t)\nonumber\\
&=&\frac{mc^2}{\sqrt{1-\omega^2(t)}}
+\frac{ee'}{c\delta'(t) [1+\omega'(t-\delta'(t))]},
\end{eqnarray}
where $\phi'(x(t),t)$ is the Lorenz-gauge scalar potential generated
by ${\rm P}'$; see (\ref{eq:4}, \ref{eq:04}) and Appendix \ref{deri}.

The change of $E$ reads:
\begin{eqnarray}
    \label{kerman}
  \Delta_{t_2|t_1} E\equiv E(t_2)-E(t_1). 
\end{eqnarray}
The kinetic energy change of ${\rm P}'$ is
\begin{eqnarray}
  \label{shah}
  \Delta_{t_2|t_1} K'\equiv K'(t_2)-K'(t_1),\\
  \label{eq:7}
K'(t)\equiv {m'c^2}/{\sqrt{1-v'^2/c^2}}.
\end{eqnarray}
We validate below that under reasonable conditions $|\Delta_{t_2|t_1}
E+\Delta_{t_2|t_1} K'|$ can be negligible, and hence the thermodynamic
work can be defined via $\Delta_{t_2|t_1}E$, or via the {\it
  gauge-invariant} $-\Delta_{t_2|t_1} K'$:
\begin{gather}
\label{eq:520}
W=-\Delta_{t_2|t_1} K', ~~ {\rm if}~~\\
  \label{eq:52} 
|\Delta_{t_2|t_1} E+\Delta_{t_2|t_1} K'|\ll |\Delta_{t_2|t_1} E|, \, 
|\Delta_{t_2|t_1} K'|.
\end{gather}
Eq.~(\ref{eq:52}) can be interpreted as an approximate conservation
law ensuring the energy transfer between ${\rm P}$ and ${\rm P'}$. The
validity of (\ref{eq:52}) confirms that $e\phi'(x,t)$ is the
time-dependent potential energy for ${\rm P}$. Defining the work via
the kinetic energy of ${\rm P'}$ is consistent the fact that this
energy can be fully transferred to heat \cite{lehrman}, e.g. by
stopping the particle by a static target.

Eqs.~(\ref{eq:520}, \ref{eq:52}) amount to the first law. Importantly,
(\ref{eq:520}, \ref{eq:52}) are written in finite differences: in the
relativistic situation the energy transfer does take a finite time,
since the energy has to pass through the EMF. Hence (\ref{eq:52})
cannot hold for a small $|t_2-t_1|$.

Note that, as implied by (\ref{eq:55}), there is only the Lorentz
force acting on ${\rm P'}$. Hence the kinetic energy differece
(\ref{shah}) can be also recovered as the (time-integrated) mechanic
work done by the Lorentz force acting on ${\rm P'}$.

Recall the non-relativistic situation, where two particles interact
directly via the Coulomb potential. There we have an exact relation
(conservation of energy)
\begin{eqnarray}
  \label{eq:56}
\frac{mv^2 (t) }{2}+\frac{ee'}{|x'(t)-x(t)|} 
=-    \frac{m'v'^2(t)}{2}+{\rm const}.
\end{eqnarray}
If (\ref{eq:57}) holds, the left-hand-side of (\ref{eq:56}) is
identified with the time-dependent Hamiltonian of ${\rm P}$, and we
get an exact correspondence between the two aspects of work; cf. the
discussion around (\ref{eq:740}). But since (\ref{eq:56}) is a
non-relativistic relation, it implies an instantaneous transfer of
energy, hence it is written as a conservation relation that holds at
any time.

The quantities in (\ref{kerman}, \ref{shah}) are calculated|in the
considered lab-frame|at the same times ($t_2$ and $t_1$,
respectively), but at different coordinates. Though events that are
simultaneous in one reference frame will not be simultaneous in
another, the relativity theory does employ such reference
frame-specific quantities, the length being the main example
\cite{gamba,str}.



Since the correct expression of the first law is open, we tried to use
instead of $E$ and $K'$ in (\ref{eq:520}, \ref{eq:52}) other
quantities, e.g. (resp.) $K(t)={mc^2}/{\sqrt{1-v^2/c^2}}$ and $E'(t)$,
where $E'(t)$ is the analogue of (\ref{eq:10}) for ${\rm P}'$
\begin{eqnarray}
E'(t)&=&\frac{m'c^2}{\sqrt{1-\omega'^2(t)}}+e'\phi(x'(t),t)\nonumber\\
&=&\frac{m'c^2}{\sqrt{1-\omega'^2(t)}}
+\frac{ee'}{c\delta(t) [1-\omega(t-\delta(t))]}.
\label{eq:11}
\end{eqnarray}
In contrast to (\ref{eq:52}), this choice (as well as several other
choices) did not lead to a sufficiently precise conservation law, i.e.
$|\Delta_{t_2|t_1} E'+\Delta_{t_2|t_1} K|$ is not negligible; see
below.


\subsection{Numerical validation}
\label{wowo_numo}

\begin{figure*}[htbp]
\centering
    \subfigure[]{
     \label{f3a}
    \includegraphics[width=0.9\columnwidth]{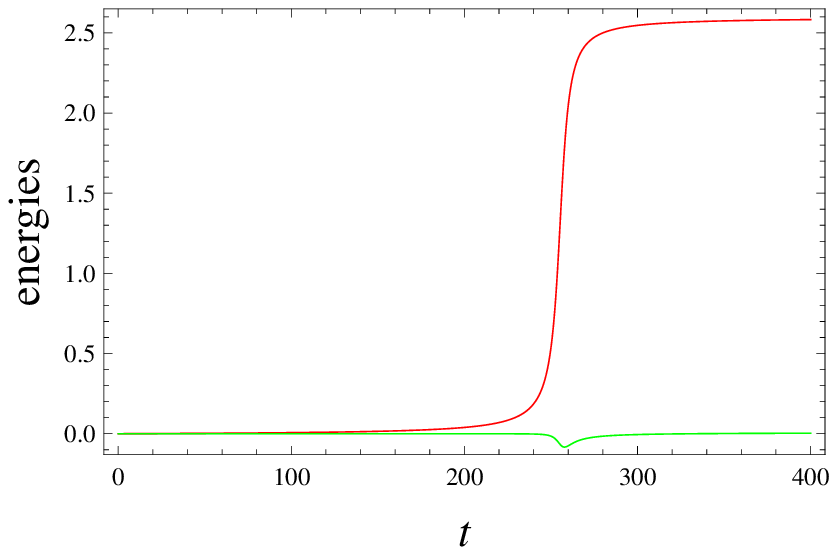} 
    }
    \subfigure[]{
     \label{f3b}
    \includegraphics[width=0.9\columnwidth]{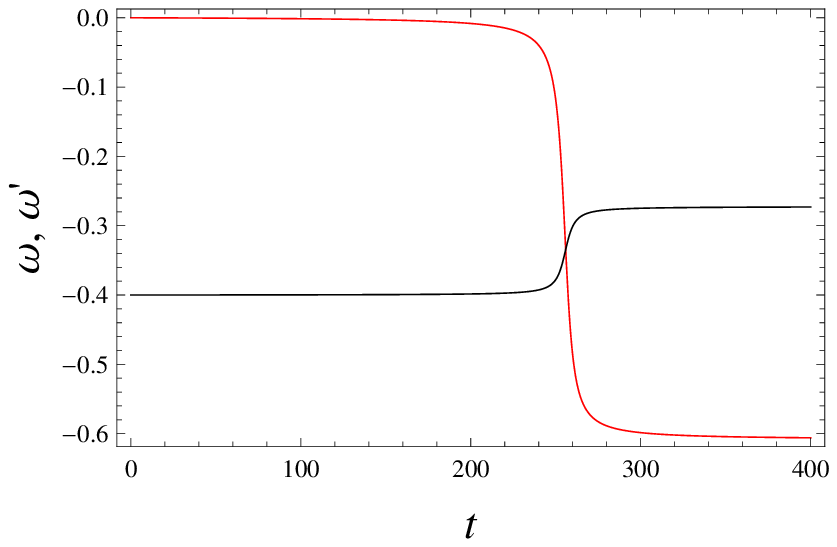} 
    }
    \caption{Repulsive motion: one particle (${\rm P}'$) falls into
      another (${\rm P}$) that is at rest initially: $ee'=1$, $c=1$,
      $l_0=100$, $m=10$, $m'=50$, $\omega_0=0$, $\omega'_0=-0.4$;
      cf. (\ref{eq:5}--\ref{eq:45}). The minimal inter-particle
      distance $x'(t)-x(t)=1.2287$ is reached at $t=255.87$. \\
      (a) $\Delta_{t|0} E$ (red) and $\Delta_{t|0}(E+K')$ (green).\\
      (b) $\omega(t)$ (red) and $\omega'(t)$ (black). }
     \label{f3}
\end{figure*}

We studied (\ref{eq:5}--\ref{eq:99}) numerically. 
Figs.~\ref{f1}--\ref{f4} show four representative examples.

Figs.~\ref{f1} refer to repelling ${\rm P}$ and ${\rm P}'$ that start
to move from a fixed distance, with zero velocities for $t\leq 0$,
i.e. $\omega_0=\omega_0'=0$ in (\ref{eq:77}). (Here the full 3d case
reduces to the considered 1d situation.)

Figs.~\ref{f2} describe a classical analogue of the annihilation
process: two attractive particles ${\rm P}$ and ${\rm P}'$ fall into
each in a finite time; their evolution again starts from a fixed
distance and with zero velocities.

Figs.~\ref{f3} show a scattering process of repelling particles: ${\rm
  P}'$ runs on ${\rm P}$ which is at rest initially.  For scattering
processes|where particles are free both initially and
finally|(\ref{eq:5}--\ref{eq:99}) predict elastic collision, if
conditions discussed around (\ref{eq:8}) hold, i.e.  the radiation
reaction can be neglected.

Figs.~\ref{f4} show a specific elastic collision, where no energy
transfers takes place between initial and final (asymptotically-free)
states, but there is a non-trivial work-exchange at intermediate
times.

Figs.~\ref{f1a}, \ref{f1b}, \ref{f2a}, \ref{f3a}, \ref{f4a} and
\ref{f4b} show that (\ref{eq:52}) holds with a good precision provided
that $t_2-t_1$ is sufficienly large. Everywhere we assume
(\ref{eq:57}): Figs. \ref{f1c}, \ref{f2b} and \ref{f3b} demonstrate
that even for modestly large values of $\frac{m'}{m}$ the motion of
${\rm P}'$ is weakly perturbed by ${\rm P}$.

The definition of work is clarified via Figs.~\ref{f1b} and \ref{f4b}:
they show that $E+K'$ is a better conserved quantity than $K'+E$;
cf. (\ref{kerman}, \ref{shah}). Hence the definition (\ref{eq:520},
\ref{eq:52}) is selected by the approximate conservation law argument.

Figs.~\ref{f1a}, \ref{f1b} and \ref{f2a} show that for a range of
initial times $E$ is strictly conserved. Hence (\ref{eq:52}) does not
hold and no work can be defined via (\ref{eq:520}). This relates to
the fact that for parameters of Figs.~\ref{f1} and \ref{f2} the
particles ${\rm P}$ and ${\rm P}'$ have zero velocities for $t<0$;
cf. (\ref{eq:77}). Due to retardation each particle sees a fixed
neighbor for some initial time. This leads to conservation of $E$ for
those times. Thus, this example illustrates the causal behavior of
work, a desirable feature ensured by the Lorenz gauge. It is absent
for the Coulomb gauge (\ref{eq:60}), where $\phi_C$ propagates
instantaneously. This example also shows that the work cannot be
defined {\it only} via (\ref{eq:520}).


Another scenario for violating (\ref{eq:52}) is seen for attracting
particles that approach each other closely; see Fig.~\ref{f2a}. Now
the inter-particle distance becomes comparable with
(\ref{eq:8}). Hence the self-force cannot be neglected, and the
considered dynamics is not applicable.


For the elastic collision displayed in Figs.~\ref{f4} the initial
overall kinetic momentum is zero $p(0)+p'(0)=0$. Hence it is zero also
finally and there is no overall energy transfer. But Fig.~\ref{f4a}
shows that the work is non-trivial at intermediate times: first the
work flows from ${\rm P}'$ to ${\rm P}$, and then goes back by the
same amount.

\section{Work in the presence of the self-force}
\label{selfself}

So far we neglected the self-force (that includes the radiation
reaction force) assuming that (\ref{eq:8}) holds.  In particular,
(\ref{eq:8}) restricts velocities of ${\rm P}$ and ${\rm P}'$. Now we
get rid of this limitation and show that the first law (\ref{eq:520},
\ref{eq:52}) still holds. Instead of (\ref{eq:5}, \ref{eq:55}) we get
from Appendix \ref{bounded} in the 1d situation
\begin{eqnarray}
  \label{gen}
 \frac{\dot \omega (t)}{[1-\omega^2(t)]^{3/2}} = 
-\frac{ee'}{mc^3}~\frac{1}{\delta'^2}~
\frac{1-\omega'(t-\delta')}{1+\omega'(t-\delta')}~~~~~~~~~~~\nonumber\\
+\frac{2e^2}{3m c^3}~
\frac{\ddot{\omega}(t)(1-\omega^2(t))
+3\omega(t)\dot{\omega}^2(t)}{[1-\omega^2(t)]^{3}},\\
\frac{\dot \omega' (t)}{[1-\omega'^2(t)]^{3/2}}  = 
\frac{ee'}{m'c^3}~\frac{1}{\delta^2}~ 
\frac{1+\omega(t-\delta)}{1-\omega(t-\delta)}~~~~~~~~~~~~~ \nonumber\\
+ \frac{2e'^2}{3m' c^3}~
\frac{\ddot{\omega}'(t)(1-\omega'^2(t))
+3\omega'(t)\dot{\omega}'^2(t)}{[1-\omega'^2(t)]^{3}},
\label{shtab}
\end{eqnarray}
where $\delta(t)$ and $\delta'(t)$ are still given by (\ref{eq:4},
\ref{eq:04}, \ref{eq:9}, \ref{eq:99}).

Technically, (\ref{gen}, \ref{shtab}) can be solved only from
the future conditions for coordinates, velocities and accelerations
\cite{kos,glass,baylis_retarded}.  This fact has to do with the
structure of the self-force. Hence we pose future conditions
\begin{gather}
  \label{gorsh}
  \left(\, x(t_{\rm
      f}), v(t_{\rm f}), \dot v(t_{\rm f}); \, x'(t_{\rm f}), v'(t_{\rm
      f}), \dot v'(t_{\rm f})\, \right), \\
\dot v(t_{\rm f})=0, \qquad \dot v'(t_{\rm f})=0,
  \label{gorsho}
\end{gather}
and numerically integrate back employing a self-consistent method; see
Appendix \ref{nunu}.  Similar methods were discussed in
\cite{travis,hsing}. The existence and uniqueness of such (Cauchy)
solutions are not generally known \footnote{ There are simple examples
  of delay-differential equations for which the forward solution
  [defined e.g. via (\ref{eq:77})] is unique, but the backward
  solution either does not exist or it is not unique
  \cite{driver_book,book_numerics,raju}.  }. There are only certain
partial results \cite{driver_back,travis,hsing,zhdanov}, e.g. that the
solution exists and it is unique for the 1D repelling case, $ee'>0$,
if the the final separation $|x(t_{\rm f})-x'(t_{\rm f})|$ is
sufficiently large \cite{kirpich}. We thus focus on this
situation. Hence the final time is so large that the particles do not
interact for $t\sim t_{\rm f}$ and for $t\sim 0$.

Figs.~\ref{f5a} and \ref{f5b} display our results for energy changes
(\ref{kerman}, \ref{shah}) obtained from solving (\ref{gen},
\ref{shtab}, \ref{eq:9}, \ref{eq:99}, \ref{gorsh}) numerically.

Fig.~\ref{f5a} shows that (\ref{eq:52}) holds, and the causality of
the energy flow is well-visible: when the energy transfer starts (at
$t\sim 200$), $\Delta_{t|0} (E+K')$ first decreases, and then
increases to a slightly smaller value: the transferred energy (work)
first goes out of ${\rm P}$|hence $\Delta_{t|0} (E+K')$ decreases|and
then it arrives at ${\rm P}$. The small mismatch between those initial
and final values is due to the energy that is radiated away. This
energy is determined from the overall kinetic energy difference
$K(0)+K'(0)-K(t_{\rm f})-K'(t_{\rm f})$ between initial and final
times, because at those times the particles are (asymptotically)
free. For parameters of Fig.~\ref{f5a} the radiated energy is small.


Fig.~\ref{f5b} presents a situation, where the velocities are
sufficiently large and the radiation is essential. The main message of
Fig.~\ref{f5b} is that the processes of radiation and energy transfer
are well-separated. This appears to be a general feature of 1d
collisions under (\ref{eq:57}). First the radiation is emitted from
${\rm P}$: $\Delta_{t|0} E$ decreases, while $\Delta_{t|0} K'=0$; see
Fig.~\ref{f5b}. No work is done at those times and (\ref{eq:52}) does
not hold. At somewhat later times, $\Delta_{t|0} E$ and $\Delta_{t|0}
K'$ start to change so that (\ref{eq:52}) holds; see
Fig.~\ref{f5b}. The work is exchanged causally: first the energy
leaves ${\rm P}$ and both $\Delta_{t|0} E$ and $\Delta_{t|0} (E+K')$
decrease. Afterwards, $\Delta_{t|0} K'$ and $\Delta_{t|0} (E+K')$
increase; see Fig.~\ref{f5b}. The validity of (\ref{eq:520},
\ref{eq:52}) is illustrated as
\begin{eqnarray}
  \label{eq:78}
W=\Delta_{t_2|t_1} K'=0.4207, ~~
\Delta_{t_2|t_1} (E+K')=3.1\times 10^{-4},\nonumber
\end{eqnarray}
where $t_1=497$ and $t_2=502.5$ in Fig.~\ref{f5b}. Hence most of the
work is exchanged in the time-interval $[t_2,t_1]$.

\section{Summary}

We started by arguing that the problem of defining and interpreting
the thermodynamic work done on a charged particle by time-dependent
electromagnetic field (EMF) is still open. In particular, the
definition of the thermodynamic work is not automatic, since the
time-dependent Hamiltonian (\ref{eq:01}) of the particle is not
gauge-invariant. Hence deeper physical reasons are needed for coming
up with a consistent definition of work. We stress that previous
attempts \cite{landau,kos,kobe,sipe,wang_pra} did not resolve this
problem. In particular, it was not clear how to formulate the first
law (that relates the work to the energy of the EMF-source), and how
to connect with the mechanic work (force times displacement). All
these issues are relevant for relativistic statistical thermodynamics
\cite{thermo_rel}.

The solution of the problem was sought along the following lines:

-- The definition of work ought to emerge from a consistent
energy-momentum tensor of the overall system (particles+EMF). In
particular, this ensures that the definition is relativistically
covariant. The standard energy-momentum tensor of EMF does not apply
to this problem, since it implies that the particle does not have a
potential energy and hence indirectly supports the choice of the
temporal gauge $\phi=0$ that leads to unacceptable conclusions for the
definition of the thermodynamic work; see section \ref{introduction}.

-- The definition should hold the first law (work-energy theorem) that
relates the energy of the work-recipient with the energy of the
work-source.

We carried out this program|within lacunae listed below|and showed
that the physically meaningful definition emerges from the Lorenz
gauge of EMF. It comes from the energy-momentum tensor (for
matter+EMF) that is proposed in section \ref{tensor}. This tensor is
gauge-invariant and holds several necessary features.  Its differences
and similarities with the standard energy-momentum tensor are
discussed in section \ref{tensor} and Appendix
\ref{standard_standard}. The thermodynamic work can be defined via the
particle's Hamiltonian in the Lorenz gauge. To an extent we were able
to check, it is only in this gauge that the thermodynamic work is
relativistically consistent and relates to the gauge-invariant kinetic
energy of the source of EMF. The latter can also be recovered as the
mechanic work done by Lorentz force acting on the source, thereby
establishing a relation between the thermodynamic and mechanic work.

Our motivation was and is to understand how to define work for
particles interacting with/via a non-stationary EMF. Besides its
obvious importance in non-equilibrium statistical mechanics,

Several questions remain open.

First, whether the first law (\ref{eq:52}) can be verified
analytically. The issue here is that delay-differential equations that
govern relativistic dynamics are notoriously difficult to study
analytically \cite{driver_book,book_numerics}. The existing
perturbative methods, which can employ (\ref{eq:57}) as a starting
point, focus either on the weakly-coupled situation \cite{synge} or on
the long-time limit \cite{bel}. Both these limits are not especially
interesting for verifying (\ref{eq:52}). An analytical derivation can
also clarify whether there are other gauges that hold the conservation
law (\ref{eq:52}) with at least the same precision as the Lorenz gauge
does.

Second, we verified the first law (\ref{eq:52}) for retarded
dynamics of two coupled charges either in the limit where the
radiation reaction is negligible (but the dynamics is still
relativistic), or when the processes of work exchange and radiation
are separated in time. (The latter is typical for 1d collisions). The
proper generalization of (\ref{eq:52}) for a situation when the work
exchange and radiation take place simultaneously is not clear yet.

Third, while the notion of work relates to the energy transfer, it
should be interesting to study the momentum transfer along the above
lines.

Fourth, our consideration is classical. We expect it to apply to the
semiclassical situation, where a non-relativistic quantum system
interacts with a classical EMF. But the quantum relativistic
situation is not clear; in this context see Ref.~\cite{silenko} for a
recent discussion of energy changes described by the Dirac's equation.

We close by mentioning several recent works which suggest that the
Lorenz gauge is fundamental for electrodynamics
\cite{russo,blondel,bobrov}, a point supported here via the definition
of the thermodynamic work.

\section*{Acknowledgements}

We thank K.V. Hovhannisyan and A. Melkikh for suggestions.
R. Kosloff, R. Uzdin and A. Levy are thanked for a seminar, where the
subject of this paper was discussed. A.E.A. was supported by COST
Action MP1209. Both authors were partially supported by the ICTP
through the OEA-AC-100. 

\begin{figure*}[htbp]
\centering
    \subfigure[]{
     \label{f1a}
    \includegraphics[width=0.9\columnwidth]{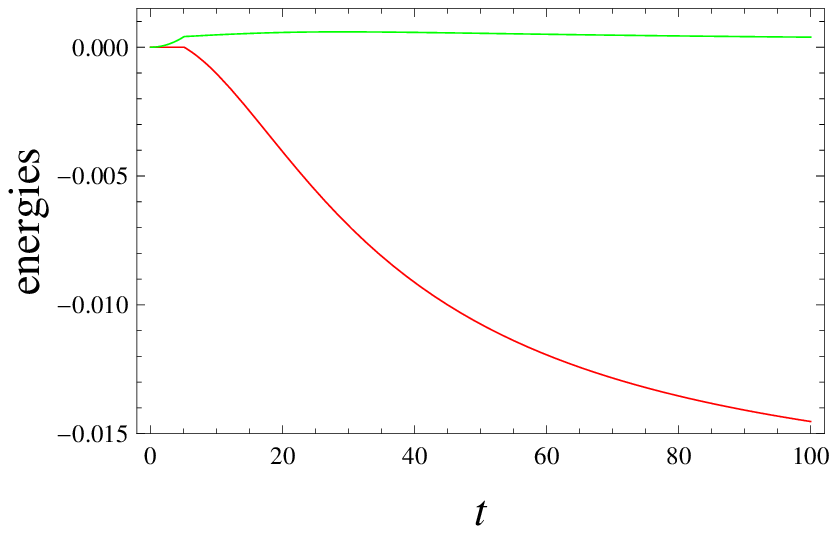} 
    }
    \subfigure[]{
     \label{f1b}
    \includegraphics[width=0.9\columnwidth]{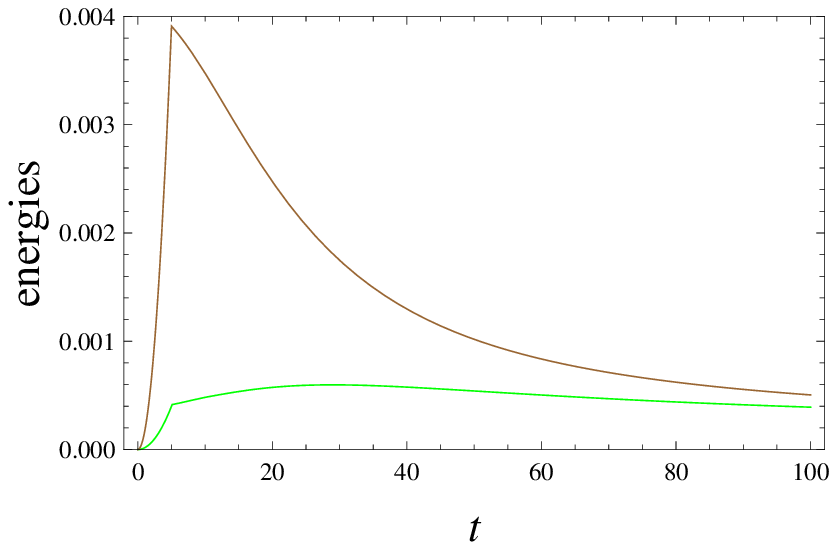} 
     }
    \subfigure[]{
     \label{f1c}
    \includegraphics[width=0.9\columnwidth]{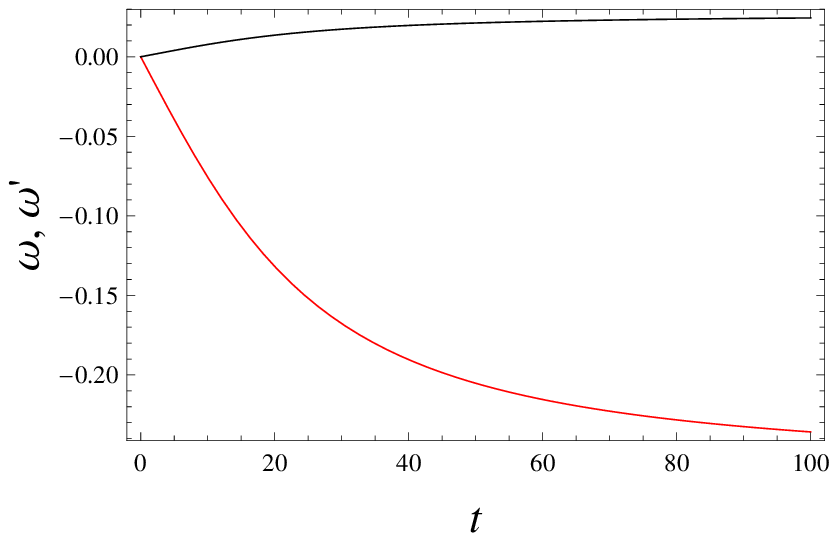} 
    }    
    \caption{Repulsive motion with initially zero velocities. The
      curves are obtained via self-consistent solution of
      (\ref{eq:5}--\ref{eq:45}) for $m=5$,
      $m'=50$, $ee'=c=1$, $l_0=5$, $\omega_0=\omega_0'=0$.\\
      (a) The energy difference $\Delta_{t|0}E$ of the light particle
      (red) and the sum $\Delta_{t|0}(E+K')$ of this energy and the
      kinetic energy $K'$ of the heavy particle (green); cf.
      (\ref{eq:520}, \ref{eq:52}).  \\
      (b) $\Delta_{t|0}(E+K')$ (green) and $\Delta_{t|0}(K+E')$
      (brown).  The former quantity is conserved better.\\
      (c) The velocity $\omega(t)$ ($\omega'(t)$) of the light (heavy)
      particle is shown by red (black) curve.  }
\label{f1}
\end{figure*}

\begin{figure*}[htbp]
\centering
    \subfigure[]{
     \label{f2a}
    \includegraphics[width=0.9\columnwidth]{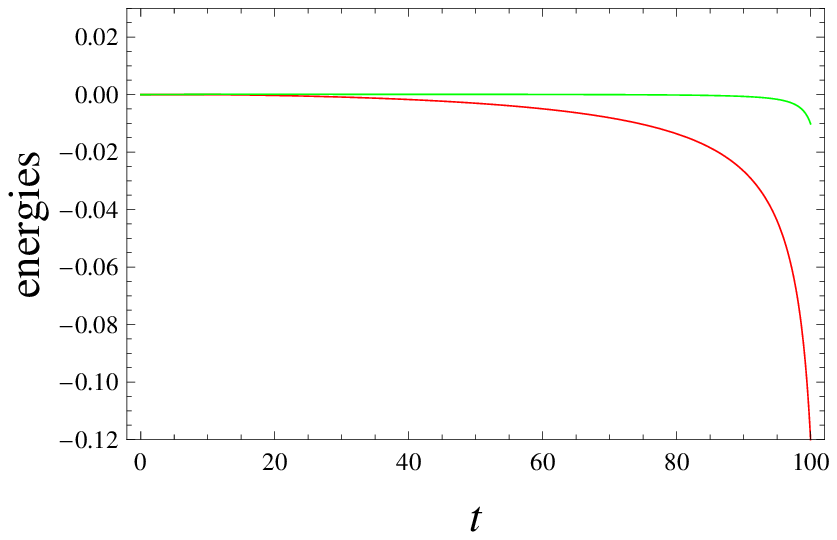} 
    }
    \subfigure[]{
     \label{f2b}
    \includegraphics[width=0.9\columnwidth]{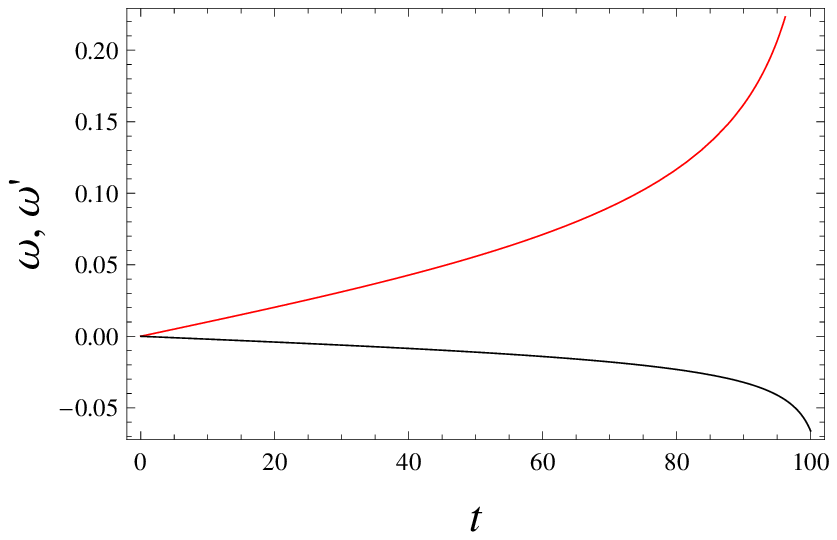} 
    }
    \caption{Attractive motion with initially zero velocities and
      $ee'=-1$, $c=1$, $l_0=10$, $m=10$, $m'=50$,
      $\omega_0=\omega'_0=0$; cf. (\ref{eq:5}--\ref{eq:45}). The
      inter-particle distance at the final time $x'(100)-x(100)=1.2219$.\\
      (a) $\Delta_{t|0}E$ (red) and $\Delta_{t|0}(E+K')$ (green).\\
      (b) $\omega(t)$ (red) and $\omega'(t)$ (black); cf.
      Figs.~\ref{f1a}--\ref{f1c}.  }
\label{f2}
\end{figure*}

\begin{figure*}[htbp]
\centering
    \subfigure[]{
     \label{f4a}
    \includegraphics[width=0.9\columnwidth]{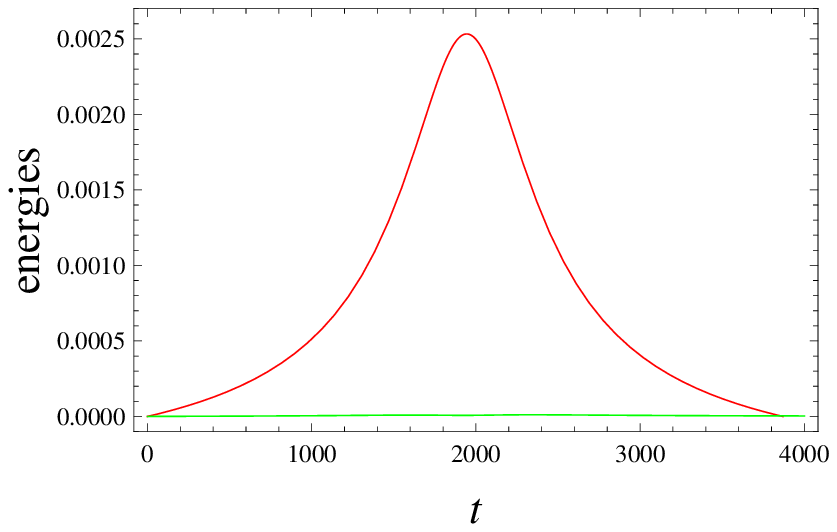} 
    }
    \subfigure[]{
     \label{f4b}
    \includegraphics[width=0.9\columnwidth]{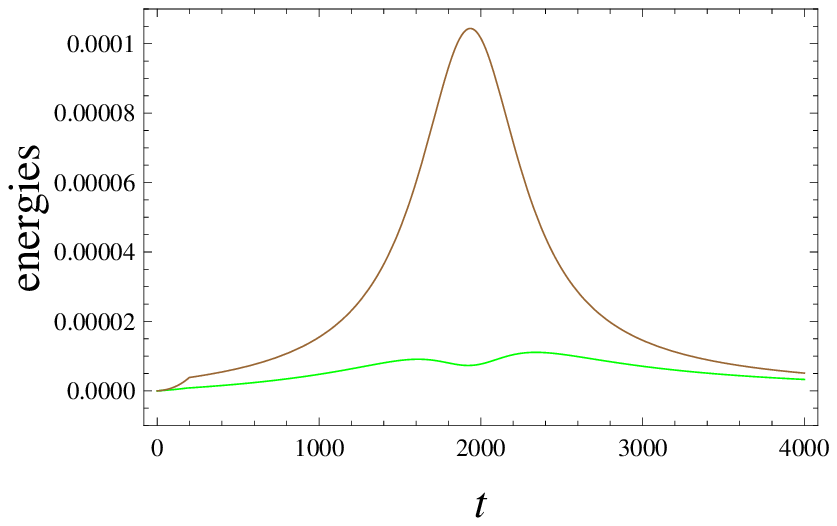} 
    }
    \caption{Scattering of particles with $ee'=1$, $c=1$, $l_0=200$,
      $m=5$, $m'=50$, $\omega_0=0.1$, $\omega'_0=-0.0100499$; cf.
      (\ref{eq:5}--\ref{eq:45}). The parameter are chosen such that
      the initial (and the final) kinetic momentum is zero:
      ${m\omega_0 }/{ \sqrt{1-\omega_0^2} } +{m'\omega_0' }/{ \sqrt{
          1-\omega_0'^2 } }=0$.  The minimal inter-particle distance
      $x'(t)-x(t)=30.57$ is reached at
      $t=1944.13$. \\
      (a) Red (upper) curve $\Delta_{t|0} E$. Green (lower) curve:
      $\Delta_{t|0}(E+K')$; cf. Fig.~\ref{f1a}. \\
      (b) Green (lower) curve: $\Delta_{t|0}(E+K')$. Brown (upper)
      curve: $\Delta_{t|0}(K+E')$.}
     \label{f4}
\end{figure*}

\begin{figure*}[htbp]
\centering
    \subfigure[]{
     \label{f5a}
    \includegraphics[width=0.9\columnwidth]{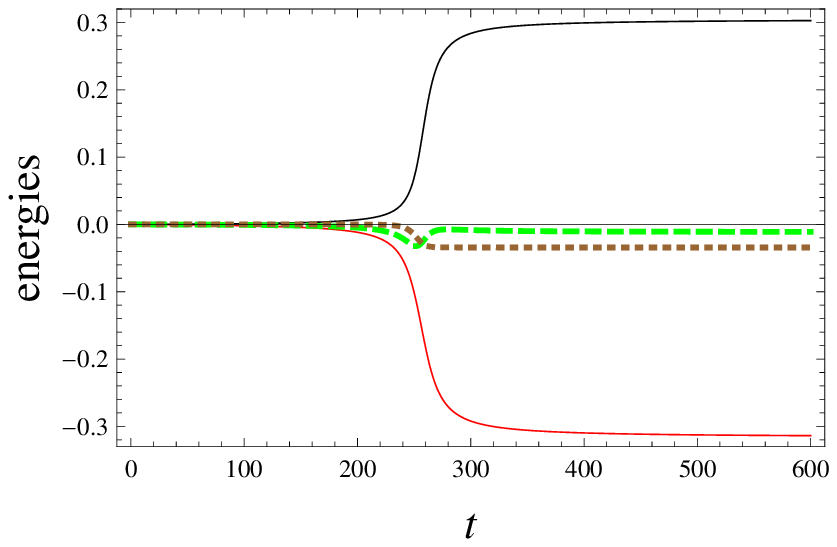} 
    }
    \subfigure[]{
     \label{f5b}
    \includegraphics[width=0.9\columnwidth]{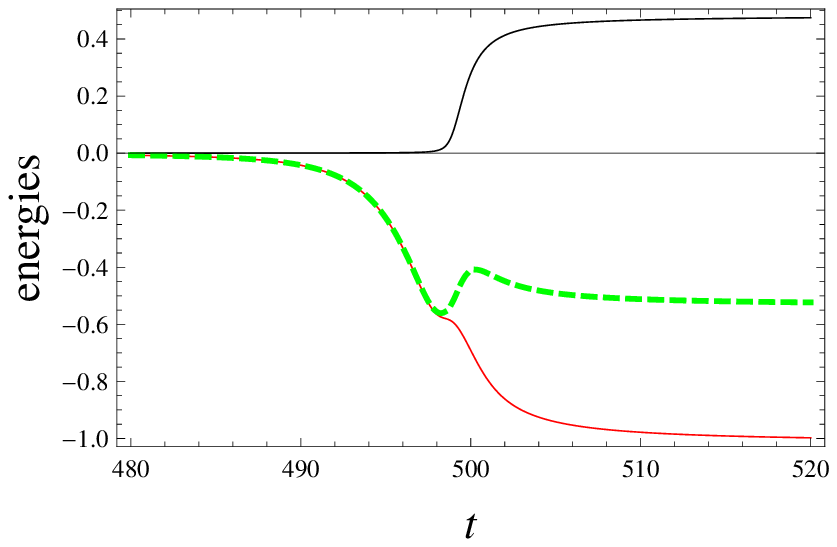} 
    }
    \caption{ Cauchy solutions of (\ref{gen}, \ref{shtab})
      for $ee'=c=1$, $m'=10$, $m'=1$. \\
      (a) Final conditions are given by (\ref{gorsh}, \ref{gorsho}):
      $t_{\rm f}=600$, $v'(t_{\rm f})=0.3$, $v(t_{\rm f})=-0.3$,
      $x'(t_{\rm f})-x(t_{\rm f})=200$. Equations of motion
      (\ref{gen}, \ref{shtab}) predict initial conditions:
      $v'(0)=0.187263$, $v(0)=0.677107$.  \\
      Red (lower) line: $\Delta_{t|0} E$. Black (upper) line:
      $\Delta_{t|0} K'$.  Green dashed line: $ \Delta_{t|0}
      (E+K')$. Brown dotted line: the sum of the Larmor's rates
      $\frac{2}{3} \int_0^t\d t\left[ \frac{\d u^k }{\d s}\frac{\d u_k
        }{\d s} + \frac{\d u'^k }{\d s'}\frac{\d u'_k }{\d s'}
      \right]\equiv -\frac{2}{3}\int_0^t\d
      \bar{t}\left[\frac{\dot{v}^2(\bar{t})}{(1-v^2(\bar{t}))^3}
        +\frac{\dot{v}'^2(\bar{t})}{(1-v'^2(\bar{t}))^3} \right]$; see
      Appendix \ref{bounded}. \\
      (b) $t_{\rm f}=700$, $v'(t_{\rm f})=0.3$, $v(t_{\rm f})=-0.7$,
      $x'(t_{\rm f})-x(t_{\rm f})=200$; $v'(0)=0.016295$,
      $v(0)=0.910459$. Red (lower) line: $ \Delta_{t|0} E$. Black
      (upper) line: $ \Delta_{t|0} K'$.  Green dashed line:
      $\Delta_{t|0} (E+K')$. Note that the energy radiation and energy
      transfer are separated from each other.  }
     \label{f5}
\end{figure*}

\appendix

\section{Gauge-freedom of non-relativistic time-dependent Hamiltonian}
\label{uno}

The Hamiltonian (\ref{eq:00}) can be deduced from a Lagrangian ${\cal
  L}(x,\dot{x};t)$ via  \cite{landau,kos}
\begin{eqnarray}
  \label{eq:61}
{\cal H}(x,\pi;t)=\pi\dot{x}-{\cal L}, \qquad 
\pi\equiv\partial_{\dot{x}} {\cal L}.  
\end{eqnarray}
The Lagrange equations stay intact if instead of ${\cal L}$ one uses
another Lagrangian $\hat{{\cal L}}$:
\begin{eqnarray}
  \label{eq:71}
\hat{{\cal L}}={\cal L}+\frac{\d \chi(x,t)}{\d t}, \qquad \frac{\d
  \chi(x,t)}{\d t}=\dot{x}\partial_x\chi+\partial_t\chi,
\end{eqnarray}
The corresponding Hamiltonian 
\begin{eqnarray}
  \label{eq:72}
\hat{{\cal
    H}}(x,\hat{\pi};t)={\cal H}(x,\pi;t)- \partial_t\chi(x,t),  
\end{eqnarray}
differs from (\ref{eq:61}) by a factor that is formally
similar to the scalar potential $\phi(\vec{x},t)$ in (\ref{eq:01}).
Hence formally the time-dependent Hamiltonian is not defined uniquely.


For the considered non-relativistic situation this non-uniqueness is
straightforward to resolve: one finds the time-independent
(non-relativistic!) Hamiltonian for the particle and the work-source
together, e.g.
\begin{eqnarray}
  \label{eq:74}
  {\cal H}_{\rm tot}(x,\pi; f,\pi_f)=  {\cal H}(x,\pi;f)+
  {\cal H}_{\rm source}(f,\pi_f),
\end{eqnarray}
where $\pi_f$ and $f$ are the canonical momentum and coordinate of the
work-source. No issue similar to (\ref{eq:72}) arises in
(\ref{eq:74}), because (\ref{eq:74}) is time-independent (one can
still add to (\ref{eq:74}) a constant). Then the physical
time-dependent Hamiltonian for the particle is found by neglecting its
backreaction onto the source: ${\cal H}(x,\pi;t)= {\cal H}(x,\pi
;f(t))$.

\comment{In particular, when the source is a heavy particle that moves almost
freely, (\ref{eq:74}) reads:
\begin{eqnarray}
  \label{eq:39}
  {\cal H}_{\rm tot}(x,\pi; X,\Pi)=  {\cal H}(x,\pi;X)+
  {\cal H}_{\rm W}(X,\Pi),
\end{eqnarray}
}

\section{Standard forms of the energy-momentum tensor }
\label{standard_standard}

Two versions of the energy-momentum tensor of electromagnetic field
(EMF) are known in literature \cite{landau,kos}
\begin{eqnarray}
  \label{beli}
&&  {\cal T}^{ik}= \frac{1}{4\pi}(-F^{il}F^k_{\,\,\,\,\,
    l}+\frac{g^{ik}}{4} F_{lm}F^{lm}), \\
  \label{nonbeli}
&&  \widetilde{\cal T}^{ik}= \frac{1}{4\pi}(-\partial^i A^l\,\,\, 
F^k_{\,\,\,\,\,
    l}+\frac{g^{ik}}{4} F_{lm}F^{lm}),\\
&& F_{ik}\equiv\partial_iA_k-\partial_kA_i.
\end{eqnarray}
$\widetilde{\cal T}^{ik}$ is deduced from the standard Lagrangian of
EMF \cite{landau}; see (\ref{eq:12}, \ref{eq:16}) in Appendix
\ref{fermi_fermi}. ${\cal T}^{ik}$ is obtained from $\widetilde{\cal
  T}^{ik}$ via the so called Belinfante method that renders a
symmetric and explicitly gauge-invariant expression
\cite{landau}. 

{\bf 1.}  First we focus on comparing (\ref{beli}) with (\ref{eq:20},
\ref{eq:166a}), because (\ref{beli}) is widely accepted as the correct
energy-momentum tensor. Then we turn to discussing (\ref{nonbeli}).

{\bf 1.1} Eq.~(\ref{beli}) does not allow to introduce potential
energy for particles [cf.~(\ref{eq:322}, \ref{eq:344})], because the
full conserved energy-momentum tensor of the EMF+matter is defined as
[cf. (\ref{eq:19})] \cite{landau}
\begin{eqnarray}
  \label{eq:58}
  {\cal T}^{ik}+\tau^{ik}, \qquad \partial_k ({\cal T}^{ik}+\tau^{ik})=0.
\end{eqnarray}
Hence according to (\ref{beli}, \ref{eq:58}) the matter has only
kinetic energy, as can be verified in detail by working out
(\ref{beli}) analogously to (\ref{eq:32}--\ref{eq:344}). This is why
(\ref{beli}) indirectly supports the choice of the temporal gauge
$\phi=0$.

{\bf 1.2} 
Recall that according to (\ref{beli}), ${\cal T}^{00}\propto
\vec{E}^2+\vec{B}^2$ is energy density, and
\begin{eqnarray}
  \label{eq:42}
c{\cal T}^{0\alpha}&=&-
\frac{c}{4\pi}(\partial^0A_\beta-\partial_\beta A^0)
(\partial^\alpha A^\beta-\partial^\beta A^\alpha)\nonumber\\
&=&\frac{c}{4\pi} \vec{E}\times\vec{B} 
\end{eqnarray}
is the energy current (Poynting vector), and ${\cal T}^{\alpha 0}$ is
the momentum density.

The Poynting vector is non-zero also for time-independent fields. This
is a known controversy in the standard definition of the EMF energy
current: stationary fields|e.g. created by a constant change and
permanent magnet, which do not require any energy cost for their
maintenance|would lead to permanent flow of energy and constant field
momentum \cite{feynman,jeffries,pugh}. In contrast, (\ref{eq:34}) is
zero for time-independent fields \cite{puthoff}. This is an advantage.

We are not aware of direct experimental results which would single out
a unique expression for EMF energy current
\cite{feynman,jeffries,slepian}. Some experiments point against the
universal applicability of the Poynting vector for the EMF energy flow
\cite{chub}.

{\bf 1.3} 
Eq.~(\ref{beli}) does not allow a clear-cut separation
between the orbital momentum and spin of EMF. Indeed, due to ${\cal
  T}^{ik}={\cal T}^{ki}$, we get for a free EMF \cite{landau}
\begin{eqnarray}
  \label{eq:48}
\partial_k {\cal M}^k_{lm}=0, \qquad
{\cal M}^k_{lm} \equiv x_m   {\cal T}_{l}^{\,\,k}-
x_l   {\cal T}_{m}^{\,\,k}.  
\end{eqnarray}
Now ${\cal M}^k_{lm}$ is conserved, but it has the form of orbital
momentum; sometimes it is also interpreted as the full angular
momentum leaving unspecified the separate contributions of orbital
momentum and spin \cite{landau} \footnote{Such quantities can be
  introduced at the level of ${\cal M}^k_{lm}$; see
  e.g. \cite{lorce_review}. But this introduction is {\it ad hoc};
  cf. (\ref{eq:25}, \ref{eq:23}, \ref{eq:28}, \ref{eq:24}).}. In
contrast, (\ref{eq:20}, \ref{eq:166a}) lead to well-defined
expressions for the angular momentum and spin that are conserved
separately for a free EMF; see Appendix \ref{angu}.  This Appendix
also explains that when EMF couples to matter only the sum of the full
orbital momentum (including that of matter) and the EMF spin is
conserved.

{\bf 1.4} Eq.~(\ref{beli}) and (\ref{eq:20}, \ref{eq:166a}) lead to
the same predictions for the (space-integrated) conserved
quantities. Indeed, using the Lorenz gauge (\ref{eq:13}) and the
equation of motion (\ref{eq:22}) for $A^i$ we get from (\ref{beli},
\ref{eq:20}):
\begin{eqnarray}
  \label{eq:47}
&&  {\cal T}^{ik}+\tau^{ik}
-\mathbb{T}^{ik}  =\frac{1}{4\pi}\partial_l B^{ikl}, \\
&&  B^{ikl}\equiv 
    A^k\partial^iA^l+A^i\partial^kA^l-A^i\partial^lA^k
    -\frac{g^{ik}}{2}\,  
    A_m\partial^mA^l.\nonumber\\
&&
\label{ho}
\end{eqnarray}
Note that (\ref{eq:47}) is not the usual freedom associated with the
choice of the energy-momentum tensor \cite{landau}. That freedom
amounts to $B^{ikl}=-B^{ilk}$, which clearly does not hold with
(\ref{ho}).

Relations $\partial_k [{\cal T}^{ik}+\tau^{ik}]=\partial_k
\mathbb{T}^{ik}=0$ amount to conservation of $\int\d^3 x\,[{\cal
  T}^{i0}+\tau^{i0}]$ and $\int\d^3 x\,\mathbb{T}^{i0}$ in time. More
generally, such a conservation law may be absent, e.g. when some
non-electromagnetic forces act on the matter. 

At any rate, we want to show that ${\cal T}^{ik}+\tau^{ik}$ and
$\mathbb{T}^{ik}$ do predict the same values for the full
(space-integrated) energy-momentum of the matter+EMF. To this end,
consider from (\ref{eq:47}, \ref{ho}) the difference of the two
predictions:
\begin{eqnarray}
  \int\d^3 x\,[{\cal T}^{i0}+\tau^{i0}] - \int\d^3
x\,\mathbb{T}^{i0} \nonumber\\
=\frac{1}{4\pi}
\partial_0 \int\d^3 x\, B^{i00}+\frac{1}{4\pi}
\int\d^3 x\, \partial_\alpha B^{i0\alpha}.
  \label{eq:67}
\end{eqnarray}
The second term in (\ref{eq:67}) contains full space-derivatives and
amounts to zero under standard boundary conditions. 
Also, $B^{i00}$ amounts to full space-derivatives,
\begin{eqnarray}
  \label{eq:68}
  B^{\alpha 00}=\frac{1}{2}\partial^\alpha[A^0A^0], ~~
  B^{0 00}=-\frac{1}{2}\partial_\alpha[A^\alpha A^0],
\end{eqnarray}
where we used the Lorenz gauge condition (\ref{eq:13}). Hence
(\ref{eq:68}) implies that $\int\d^3 x\, B^{i00}=0$. Thus, we get from
(\ref{eq:67})
\begin{eqnarray}
  \label{eq:6767}
  \int\d^3 x\,[{\cal T}^{i0}+\tau^{i0}] = \int\d^3
x\,\mathbb{T}^{i0}. 
\end{eqnarray}

{\bf 1.5} 
Eq.~(\ref{beli}) predicts a non-negative expression 
\begin{eqnarray}
  \label{eq:44}
  {\cal T}^{00}=\frac{1}{8\pi}( \vec{E}^2+\vec{B}^2  )
\end{eqnarray}
for the energy density of EMF \cite{landau}. In contrast, according to
(\ref{eq:20}) the energy density of EMF is generally not positive. It
is positive for stationary fields, and it is positive for a radiation
emitted by a point particle, where (\ref{eq:20}) and (\ref{beli})
agree with each other; see Appendix \ref{radiation}. The
non-positivity should not be regarded as a drawback, e.g. because once
for point particles the diverging terms for (\ref{beli}) are
renormalized away, the energy density of EMF is not anymore strictly
positive.

{\bf 1.6} 
Another difference between (\ref{beli}) and (\ref{eq:20})
for a free EMF is that the zero-trace relation ${\cal T}^{i}_i=0$ is
always true, while ${T}^{i}_i$ is generally not zero;
cf.~(\ref{eq:166a})]. Now ${\cal T}^{i}_i=0$ is generally related to
the zero mass of photon. For photons we also get that ${T}^{i}_i=0$;
see Appendix \ref{radiation}.  However, it is not generally true that
a superposition of two or more photons (e.g. bi-photon) has a
zero-mass; see e.g. \cite{gabovich}. Physically, this means that we
should not expect the zero-trace relation for an arbitrary EMF.

{\bf 2.} We now turn to discussing the features of (\ref{nonbeli}).

{\bf 2.1} Eq.~(\ref{nonbeli}) is non-symmetric even for the free
EMF. Hence the desired relation between the energy current and
momentum density is generally violated: $\widetilde{\cal
  T}^{0\alpha}\not=\widetilde{\cal T}^{\alpha 0}$. This is not
physical.

{\bf 2.2} 
The symmetry of (\ref{nonbeli}) for a free EMF also means
that if one introduces the orbital momentum as $\widetilde{{\cal
    M}}^k_{lm} \equiv x_m \widetilde{{\cal T}}_{l}^{\,\,k}- x_l
\widetilde{{\cal T}}_{m}^{\,\,k}$, it is generally not conserved:
$\partial_k \widetilde{{\cal M}}^k_{lm}\not=0 $. This is not physical.

{\bf 2.3} Eq.~(\ref{nonbeli}) is neither explicitly gauge-invariant,
nor it allows to single out any specific gauge \footnote{This point
  can be reformulated as follows \cite{lorce_review}. It is based on
  separating the full potential $A_i$ into a physical
  (i.e. gauge-invariant) part and the pure gauge: $A_i=A_i^{\rm
    phys}+A_i^{\rm pure}$, where $A_i^{\rm pure}=\partial_i \chi$
  \cite{lorce_review}. Expectedly, the modified (i.e. gauge-invariant)
  analogue of (\ref{nonbeli}) is given by the same expression, where
  $A^l\to A^{l~{\rm phys}}$. The modified expression is now
  gauge-invariant, but is still not unique, because now there is a freedom
  in choosing $A_i^{\rm phys}$. }.

{\bf 2.4} 
Eq.~(\ref{nonbeli}) relates to the following energy-momentum
tensor for EMF+matter:
\begin{eqnarray}
  \label{eq:59}
 \partial_k ( \widetilde{\cal T}^{ik}+\tau^{ik}+ \frac{1}{c}A^iJ^k)=0.
\end{eqnarray}
This follows from $\widetilde{\cal T}^{ik}={\cal T}^{ik}
-\frac{1}{4\pi}\partial^lA^i \, F^{k}_{\,\,\, l} $ [see (\ref{beli},
\ref{nonbeli})] and from (\ref{eq:58}). Comparing (\ref{eq:59}) with
(\ref{eq:20}), we see that (\ref{eq:59}) predicts analogues of
(\ref{eq:322}, \ref{eq:344}), but without singling out the Lorenz
gauge (\ref{eq:13}).

\section{Fermi's Lagrangian for free EMF}
\label{fermi_fermi}

The purpose of this Appendix is to derive (\ref{eq:166a}) from the
Fermi's Lagrangian for a free classical electro-magnetic field
(EMF) \cite{bogo,fermi,oost}. The standard Lagrangian reads
\begin{eqnarray}
  \label{eq:12}
{\cal L}=-\frac{1}{16\pi} F_{ik}F^{ik}, \qquad 
F_{ik}\equiv\partial_iA_{k}-\partial_kA_{i}.
\end{eqnarray}
where $A^i=(\phi,\vec{A})$ is the 4-potential.

For the Lorenz gauge (\ref{eq:13}), an alternative Lagrangian was
proposed by Fermi \cite{fermi,bogo,oost}. It is obtained from
(\ref{eq:12}) upon using (\ref{eq:13}) and neglecting full derivatives
\begin{eqnarray}
  \label{eq:14}
  L=-\frac{1}{8\pi}\partial_iA_k\partial^i A^k.
\end{eqnarray}
The equations of motion $\partial_k \frac{\partial L}{\partial
  [\partial_k A^i]}=\frac{\partial L}{\partial
  A^i}$
are
\begin{eqnarray}
  \label{eq:15}
  \partial_k\partial^k A_i=0,
\end{eqnarray}
which are consistent with the Lorenz gauge (\ref{eq:13}). The latter
is to be considered as a condition imposed on (\ref{eq:14}).


Given (\ref{eq:14}) and recalling the standard expression for the the
energy-momentum tensor \cite{landau,bogo}
\begin{eqnarray}
  \label{eq:16}
  T_{i}^{\,\,k}=\partial_i A_l\, 
  \frac{\partial  L}{\partial [\partial_k A_l]}
-{\cal L}\delta_{i}^k,
\end{eqnarray}
we obtain from (\ref{eq:14}):
\begin{eqnarray}
   T_{i}^{\,\,k}
=-\frac{1}{4\pi}\, \partial_iA_l\, \partial^k A^l +\frac{1}{8\pi}\,
\delta_{i}^k \,
\partial_nA_m\,\partial^n A^m,
  \label{eq:166}
\end{eqnarray}
where $\delta_{i}^k$ is the Kroenecker delta. Eqs.~(\ref{eq:166},
\ref{eq:15}) show that the tensor is symmetric
and holds energy-momentum conservation
\begin{eqnarray}
  \label{eq:18}
&&  T_{ik}=T_{ki}, \\
  \label{eq:17}
&&  \partial_k   T_{i}^{\,\,k}=0.
\end{eqnarray}

The invariance of a Lagrangian under rotations implies the following
general relation between the energy-momentum tensor and angular
momentum tensor \cite{bogo}:
\begin{eqnarray}
  \label{eq:28}
M^k_{lm}&=&  x_m   {T}_{l}^{\,\,k}- x_l   {T}_{m}^{\,\,k}+  S^k_{lm},\\
  \label{eq:244}
S^k_{lm}&=& \frac{\partial L}{\partial
    [\,\partial_k A^m\,]}\, A_l - \frac{\partial  L}{\partial
    [\,\partial_k A^l\,]}\, A_m,
\end{eqnarray}
where ${T}_{l}^{\,\,k}$ is given by (\ref{eq:16}),
$S^k_{lm}$ is the internal angular momentum (spin), while
$x_m   {T}_{l}^{\,\,k}- x_l   {T}_{m}^{\,\,k}$
is the orbital momentum. Using (\ref{eq:14})
we obtain for the spin tensor \cite{bogo,oost}
\begin{eqnarray}
S^k_{lm}
=-\frac{1}{4\pi}\left(
A_l\, \partial^k A_m -A_m\,\partial^k A_l
\right).
  \label{eq:24}
\end{eqnarray}
Employing (\ref{eq:15}, \ref{eq:18}, \ref{eq:17}) we get that the
angular momentum and spin tensor are conserved separately
\cite{bogo,oost}:
\begin{eqnarray}
  \label{eq:29}
\partial_k  M^k_{lm}=\partial_k S^k_{lm}=0,
\end{eqnarray}
as should be for a free field. The symmetry (\ref{eq:18}) is
crucial for the existence of two separate conservation laws
(\ref{eq:29}).

The standard approaches to the energy-momentum tensor of EMF are
recalled in Appendix \ref{standard_standard}. The spin and orbital
momentum of EMF is reviewed in \cite{lorce_review,babi}.

\section{Energy momentum-tensor for free radiation}
\label{radiation}

Here we discuss the energy-momentum tensor (\ref{eq:166a}) for
free radiation. Within this appendix we put $c=1$.

Consider a charge $e$ moving on a world-line $z^i(s)$. Define $u^i=\d
z^i(s)/\d s$ for the 4-velocity. Recall that $u^iu_i=1$. Let $w^i$
and $c^i$ be defined as follows
\begin{eqnarray}
  \label{eq:62}
&&  w^iw_i=-1, \qquad u^iw_i=0, \\
&& c^i=u^i+w^i, \qquad c^ic_i=0. 
  \label{eq:622}
\end{eqnarray}
Given an observation event with 4-coordinates $x^i$, there is a unique
point $z^i(s_{\rm ret})$, so that the light signal emitted from a
$z^i(s_{\rm ret})$ reached $x^i$. Define using (\ref{eq:62},
\ref{eq:622}) \cite{kos}:
\begin{eqnarray}
  \label{eq:40}
&& R^i=x^i-z^i(s_{\rm ret}), \qquad R^iR_i=0, \\
&& R^i=\rho c^i, \qquad \rho =R^iu_i. 
  \label{eq:400}
\end{eqnarray}
When $x^i$ changes, $z^i(s_{\rm ret})$ changes as well. Hence there is
a problem of calculating derivatives \cite{kos}. There are 3 main
formulas here (recall that $\partial_i\equiv \partial/\partial x^i$):
\begin{eqnarray}
  \label{eq:63}
&&  \partial_k s_{\rm ret}=c_k, \\
  \label{eq:64}
&& \partial_k R^i=\delta^i_k-u^ic_k, \\
  \label{eq:65}
&& \partial_k \rho=u_k+c_k(R_ia^i-1), \qquad a^i\equiv\d^2 x^i(s)/\d
s^2. ~~~~~~
\end{eqnarray}
To derive (\ref{eq:63}, \ref{eq:64}), note from (\ref{eq:40}):
$ \partial_k R^i=\delta^i_k-u^i\partial_k s$. This relation together
with $R^iR_i=0$ implies: $R_i\partial_k R^i=0=R_k-R^iu_i\partial_k
s$. Together with (\ref{eq:400}) this leads to (\ref{eq:63}) and then
to (\ref{eq:64}). Eq.~(\ref{eq:65}) is deduced from $\partial_k
\rho=\partial_k (R^iu_i)$ using $\partial_k u^i=a^i\partial_k
s=a^ic_k$. 

The Lienard-Wichert potential of a charge $e$ reads
\begin{eqnarray}
  \label{eq:6464}
  A^i=eu^i/\rho.
\end{eqnarray}
Employing (\ref{eq:63}--\ref{eq:65}) we obtain
\begin{eqnarray}
\label{eq:0066}
F^{ik}&=&  \partial^i A^k-\partial^k A^i = e(R^i\omega^k-R^k\omega^i ), \\
\omega^i&=&\frac{a^i}{\rho^2}+\frac{u^i(1-a_lR^l)}{\rho^3},\\
  \label{eq:66}
  \partial^i A^l\partial^k A_l&=&
\frac{e^2c^ic^k [a_la^l+(a_lw^l)^2]}{\rho^2}\\
  \label{eq:666}
&-&(a_lw^l)\frac{e^2[c^iw^k+c^kw^i]}{\rho^3}
+\frac{e^2w^iw^k }{\rho^4}, ~~~~~\\
  \label{eq:6666}
  \partial_k A^l\partial^k A_l&=& (a_lw^l)\frac{2e^2}{\rho^3}
-\frac{e^2}{\rho^4}.
\end{eqnarray}
These expressions determine from (\ref{eq:166a}) the energy-momentum
tensor of EMF.

Note that the right-hand-side of (\ref{eq:66}) is the only term that
scales as $\rho^{-2}$. This is the term responsible for the
energy-momentum of the emitted radiation. It coincides with the
radiation energy-momentum tensor obtained from the standard expression
(\ref{beli}) \cite{kos}. In particular, it is symmetric and has zero
trace due to (\ref{eq:622}).

\section{Angular momentum}
\label{angu}

Here we shall connect the energy-momentum tensor $\mathbb{T}^{ik}$ to
the angular momentum tensor. 

It is seen from (\ref{eq:20}) that $\mathbb{T}^{ik}$ is not symmetric:
$\mathbb{T}^{ik}\not=\mathbb{T}^{ki}$.  This asymmetry has a physical
meaning and it relates to the spin of EMF \footnote{A non-symmetric
  energy-momentum tensor implies a generalized gravity; see, e.g.
  \cite{trautman,sciama,bregman} for examples of such
  theories. }. Recall the following general relation between the
orbital momentum tensor $\mathbb{O}^k_{lm}$ and energy-momentum tensor
$\mathbb{T}^{ik}$ \cite{landau,bogo}
\begin{eqnarray}
  \label{eq:23}
  \mathbb{O}^k_{lm}= x_m   \mathbb{T}_{l}^{\,\,k}- x_l
  \mathbb{T}_{m}^{\,\,\, k}.
\end{eqnarray}
The orbital momentum of matter is already included into
$\mathbb{O}^k_{lm}$. Due to $\mathbb{T}^{ik}\not=\mathbb{T}^{ki}$, the
orbital momentum is not conserved:
$\partial_k\mathbb{O}^k_{lm}\not=0$. This is natural, since it is the
full angular momentum $\mathbb{M}^k_{lm}=\mathbb{O}^k_{lm}+S^k_{lm}$
(orbital+spin of EMF) that should be conserved. We can thus deduce the spin
tensor $S^k_{lm}$ from the conservation law
\begin{eqnarray}
  \label{eq:25}
  \partial_k \mathbb{M}^k_{lm}=  
\partial_k\left (\mathbb{O}^k_{lm}+S^k_{lm}\right) =0. 
\end{eqnarray}
Eqs.~(\ref{eq:46}--\ref{eq:25}) and the fact that $S^k_{lm}$ should
be a quadratic function of $A_i$ imply
\begin{eqnarray}
S^k_{lm}
=-\frac{1}{4\pi}\left(
A_l\, \partial^k A_m -A_m\,\partial^k A_l
\right).
  \label{eq:24a}
\end{eqnarray}
This expression has formally the same shape as the spin tensor derived
in \cite{bogo,oost} for a free EMF via the Fermi's Lagrangian; see
Appendix \ref{fermi_fermi}. 

Hence the matter-field coupling leads (as expected) to exchange
(\ref{eq:25}) between the orbital momentum and the spin. If this
coupling is absent, then $\mathbb{T}^{ik}$ is symmetric; hence
$\mathbb{O}^k_{lm}$ and $S^k_{lm}$ are conserved separately.  These
two points|conservation of $\mathbb{O}^k_{lm}+S^k_{lm}$ under
matter-EMF coupling and separate conservation of $\mathbb{O}^k_{lm}$
and $S^k_{lm}$ for free EMF|are specific features of
(\ref{eq:25}--\ref{eq:24a}) that distinguish it from other proposals
for angular momentum of EMF; see \cite{lorce_review,babi} for a review of
those proposals.

\section{Derivation of (\ref{eq:5}--\ref{eq:99})}
\label{deri}

Consider two interacting point particles ${\rm P}$ and ${\rm
P}'$; we denote their parameters by primed and unprimed letters. 
In (\ref{eq:46}, \ref{eq:31}), the
current $J^k$ divides into two contributions, each of them is
conserved separately
\begin{eqnarray}
  \label{eq:37} J^k=j^k+j'^k, ~~~
\partial_k j^k=\partial_k j'^k=0.
\end{eqnarray} 
The EMF field $A^k$ in (\ref{eq:22}, \ref{eq:31})
also divides into two parts:
\begin{eqnarray}
  \label{eq:38} A^k=\A^k+\A'^k, ~~~
\partial_k\A^k=\partial_k\A'^k=0, \\
\partial_i\partial^i\A^k=\frac{4\pi}{c} j^k, ~~
\partial_i\partial^i\A'^k=\frac{4\pi}{c}j'^k.
  \label{eq:388}
\end{eqnarray} Hence $\A^k$ ($\A'^k$) is created by $j^k$ ($j'^k$).

Equations of motion for ${\rm P}$ and ${\rm P}'$ are deduced from
(\ref{eq:31}) noting that the self-interaction is neglected and the
point-particle limit is taken; cf. (\ref{eq:36}, \ref{eq:366}). 
These equations read \cite{synge,driver,
driver_existence,stabler,cornish,leiter,beil,franklin,kirpich}:
\begin{eqnarray}
  \label{eq:49} mc^2 \frac{\d u^k}{\d s}=eu_l f'^{kl}, ~~ m'c^2
\frac{\d u'^k}{\d s'}=e'u'_l f^{kl},
\end{eqnarray} where 
\begin{gather}
f^{kl}=\partial^k a^l-\partial^l a^k, \quad
f'^{kl}=\partial^k a'^l-\partial^l a'^k,\\ 
\d s=c\d t\sqrt{1-v^2/c^2}, \quad \d s'=c\d t\sqrt{1-v'^2/c^2}.
\end{gather}
Thus the following (self-interaction-excluded) energy-momentum tensor
is conserved [cf.~(\ref{eq:20})]
\begin{gather}
\partial_k \widetilde{\mathbb{T}}^{ik}=0,\\
  \label{eq:41} \widetilde{\mathbb{T}}^{ik} =
-\frac{1}{4\pi}[\,\partial^i \A_l \,
\partial^k \A'^l + \partial^i \A'_l \,
\partial^k \A^l -g^{ik}\partial_n \A_m \partial^n \A'^m ] \nonumber\\
+ \tau^{\, ik}+ \tau'^{\, ik}+\frac{1}{c}\A^i j'^k +\frac{1}{c}\A'^i
j^k,
  \label{eq:411}
\end{gather}
where $\tau^{\, ik}$ and $\tau'^{\, ik}$ are the
energy-momentum tensors of ${\rm P}$ and ${\rm P}'$; see
(\ref{eq:19}).

As usual, we shall select the retarded (Li\'enard-Wiechert) solutions
of (\ref{eq:388}); see \cite{landau,kos} and Appendix
\ref{radiation}. In contrast to ${\mathbb{T}}^{ik}$ that diverges in
the point-particle limit, $\widetilde{\mathbb{T}}^{ik}$ is already a
convergent tensor, i.e. the energy of EMF and particles can be
calculated via $\int\d ^3 x\, \widetilde{\mathbb{T}}^{00}(t,\vec{x})
$.

We focus on the 1D situation, where the particles ${\rm P}$ and ${\rm
  P}'$ move on a line; see (\ref{eq:0}). The retarded solutions for
$\A^i =(\phi,A)$ and $\A'^i=(\phi',A')$ in (\ref{eq:388}) read
\cite{landau}:
\begin{eqnarray}
  \label{eq:011}
  \phi(x',t)&=&\frac{e}{[x'-x(t-\delta)][1-\omega(t-\delta)]},  \\
  A(x',t)&=&\frac{e\omega(t-\delta)   }
  {[x'-x(t-\delta)][1-\omega(t-\delta)]},  \\
  \phi'(x,t)&=&\frac{e'}{[x'(t-\delta')-x][1+\omega'(t-\delta')]},  \\
  A'(x,t)&=&\frac{e'\omega'(t-\delta')   }
  {[x'(t-\delta')-x][1+\omega'(t-\delta')]},
  \label{eq:0111}
\end{eqnarray}
where the delays $\delta(t)$ and $\delta'(t)$ hold
\begin{eqnarray}
  \label{venk}
  c\delta(t) = x'-x(t-\delta(t)), \\
  c\delta'(t) = x'(t-\delta'(t))-x.
  \label{venk1}
\end{eqnarray}

To derive the equations of motion from (\ref{eq:49},
\ref{eq:011}--\ref{venk1}) recall that $f^{kl}$ [$f'^{kl}$] in
(\ref{eq:49}) is taken at $x'=x'(t)$ [$x=x(t)$]:
\begin{eqnarray}
  \label{eq:1}
  \dot p(t) &=& 
-ee' \,\frac{1-v'(t-\delta')/c
  }{1+v'(t-\delta')/c} ~ \frac{1}{[x(t)-x'(t-\delta')]^2},~~~~~~ \\
  \label{eq:2}
  \dot p'(t) 
&=&ee' \,\frac{1+v(t-\delta)/c
  }{1-v(t-\delta)/c} ~ \frac{1}{[x'(t)-x(t-\delta)]^2},
\end{eqnarray}
where 
\begin{eqnarray}
  \label{eq:3}
&& p={mv}/{\sqrt{1-v^2/c^2}}, ~~~   
p'={mv'}/{\sqrt{1-v'^2/c^2}}, \\
&& v(t)=\dot x(t)\equiv c~\omega(t), \qquad   v'(t)=\dot x'(t)\equiv
c~\omega'(t).~~~~~~~~
  \label{eq:6}
\end{eqnarray}
We get (\ref{eq:5}--\ref{eq:99})
from (\ref{venk}--\ref{eq:6}).

\section{Solving self-consistently delay-differential equations }
\label{nunu}

{\bf 1.}
Let us explain how to solve (\ref{eq:5}--\ref{eq:77}).
The method described below was first suggested in \cite{synge}; see
also \cite{franklin} for a recent discussion.

We start with an initial function $\omega_0'(t)$ that holds
$\omega_0'(t\leq 0)=\omega_0'$; cf.~(\ref{eq:77}). Then (\ref{eq:5},
\ref{eq:9}) become ordinary differential equations for $\omega(t)$ and
$\delta'(t)$. They are solved for $t>0$ with initial conditions
$\omega(0)=\omega_0$ and $\delta'(0)$ from (\ref{eq:45});
cf. (\ref{eq:77}). The solution is denoted by $\omega_0(t)$. This
function is extended to $t<0$ via (\ref{eq:77}):
$\omega_0(t<0)=\omega_0$.

Next, $\omega_0(t)$ is put into (\ref{eq:55}, \ref{eq:99}), and these
equations are solved for $t>0$, $\omega'(0)=\omega_0'$ and $\delta(0)$
from (\ref{eq:45}).  The solution $\omega_1'(t)$ is again extended to
$t<0$ via (\ref{eq:77}): $\omega_1'(t<0)=\omega_0'$.
Iterations are continued till convergence.

{\bf 2.}
We turn to solving (\ref{gen}, \ref{shtab}, \ref{eq:9},
\ref{eq:99}) given (\ref{gorsh}, \ref{gorsho}).

Now the initial function $x'(t)=x_0'(t)$ is defined for $t<t_{\rm f}$ so
that it holds (\ref{gorsh}, \ref{gorsho}). For solving
(\ref{gen}, \ref{eq:9}) we need to know $\delta'(t_{\rm f})$. It is
found from (\ref{eq:04}), i.e. from 
\begin{eqnarray}
  \label{eq:26}
c\delta_0'(t_{\rm f})=x_0'(t_{\rm f}-\delta'(t_{\rm f}))-x(t_{\rm f}).  
\end{eqnarray}
Hence this initial condition will change from iteration to
another. Now (\ref{gen}, \ref{eq:9}) can be solved as ordinary
differential equations backward in $t$ from $t_{\rm f}$ to some
$T_{\rm i}\ll t_{\rm f}$.  Then the solution is continued to $t<T_i$
by assuming that $\omega(t)=\omega(T_i)$ for $t<T_i$. This assumption
is needed, because it is impossible to integrate numerically from
$t_{\rm f}$ till $-\infty$. Effectively, this means that the particles
did not interact in the remote past; or (alternatively) that they
interacted so strongly that $\omega(t)=v(t)/c \simeq 1$ for $t<T_i$.

Overall, the solution defines $x_0(t)$ with which we repeat the above
step for (\ref{shtab}, \ref{eq:99}), e.g. $x_0(t)$ is put into
(\ref{shtab}, \ref{eq:99}), and now we have instead of (\ref{eq:26}):
$c\delta_0(t_{\rm f})=x'(t_{\rm f}) -x_0(t_{\rm f}-\delta_0(t_{\rm
  f}))$. The backward solution of ordinary-differential (\ref{shtab},
\ref{eq:99}) is continued as $\omega'(t)=\omega'(T_i)$ for $t<T_i$.
Once iterations converged, we assure by direct replacement that
(\ref{gen}, \ref{shtab}, \ref{eq:9}, \ref{eq:99}) do hold for $T_i\ll
t<\tau$.

\comment{
Let (\ref{eq:5}--\ref{eq:99}) are solved via (\ref{eq:39},
\ref{eq:399}) from $t=0$ to $t=\tau>0$. Now (\ref{eq:50}) are taken
from that solution and (\ref{eq:5}--\ref{eq:99}) are solved back from
$\tau$ to $t=0$ producing $z^T(2\tau)$; cf. (\ref{eq:50}). Even when
this solution exists (and when it is unique) one should not
generally expect $z(\tau)=z^T(2\tau)$, e.g. because due to
(\ref{eq:51}), $z(t)$ solves the equations of motion only for $t\geq
0$, while $z^T(2\tau)$ does not have this constraint. Nevertheless,
\begin{eqnarray}
  \label{eq:54}
z(0)\approx z^T(2\tau),
\end{eqnarray}
can hold in many cases of practical interest, also including cases
with relativistic velocities. Whenever $z(\tau)$ is sufficiently far
from $z^T(2\tau)$, we can speak on the effective arrow of
time. Figs.~\ref{f1c} and \ref{f2c} provide examples on this. For
initial conditions of Fig.~\ref{f1c} [full curves] the dynamics is
approximately reversible (with absolute precision $\simeq 10^{-4}$);
cf. (\ref{eq:54}).  In contrast, dotted curves in Fig.~\ref{f1c}
correspond to initial conditions, where the dynamics is not
reversible, i.e. (\ref{eq:54}) does not hold: for the light particle
$|x(0)-x^{\rm T}(2\tau)|\simeq 1$.
}

\section{Self-force}
\label{bounded}

The electrodynamic self-force adds to (\ref{eq:49})
\cite{landau,kos,glass}:
\begin{eqnarray}
\label{bori}
  m\frac{\d u^k }{\d s}=
\frac{e}{c^2}u_l\, f'^{kl}+\frac{2e^2}{3c^2}
[\,\frac{\d^2 u^k }{\d s^2}+\frac{\d u^l }{\d s}\frac{\d u_l}{\d s}u^k], \\
  m'\frac{\d u'^i}{\d s'}=
\frac{e'}{c^2}u'_l\, f^{kl}+\frac{2e'^2}{3c^2}
[\,\frac{\d^2 u'^k }{\d s'^2}+\frac{\d u'^l }{\d s'}\frac{d u'_l}{\d s'}u'^k].
\label{boribori}
\end{eqnarray}
Constraints $u_k\frac{\d u^k }{\d s}=u'_k\frac{\d u'^k }{\d s}=0$ in
(\ref{bori}, \ref{boribori}) are ensured due to $u_ku^k=u'_ku'^k=1$
and $f'^{kl}=-f'^{lk}$, $f^{kl}=-f^{lk}$;
cf. (\ref{eq:33}). 
Eqs.~(\ref{bori}, \ref{boribori}) lead to (\ref{gen}, \ref{shtab}) via
(\ref{eq:011}--\ref{venk1}).

Note the following interpretation of (\ref{bori}) \cite{kos,glass}:
\begin{eqnarray}
\label{borik}
  \frac{\d }{\d s}[mu^k -\frac{2e^2}{3c^2}\frac{\d u^k }{\d s}  ]=
 \frac{2e^2}{3c^2} u^i \frac{\d u^k }{\d s}\frac{\d u_k }{\d s}
+\frac{e}{c^2}u_l\, f'^{kl}, ~~~\\
\frac{1}{c^2}\frac{\d u^k }{\d s}\frac{\d u_k }{\d s}=-
\frac{(\vec{\omega}\dot{\vec{\omega}}) ^2+(1-\omega^2)\dot{\vec{\omega}}^2}
{(1-\omega^2)^3}\leq 0, \quad \vec{\omega}=\frac{\vec{v}}{c}.~~
\end{eqnarray}
It was suggested that $ \frac{2e^2}{3c^2} u^i \frac{\d u^k }{\d
  s}\frac{\d u_k }{\d s}$ can be related to the emitted (radiated
away) 4-momentum \cite{kos,glass} (Larmor's rates), while
$-\frac{2e^2}{3c^2}\frac{\d u^k }{\d s}$ is to be related with the
bound 4-momentum, i.e. the momentum of an effective ``cloud'' around
the particle \cite{kos,glass}.

We did not find a serious support for these interpretations for the
two-particle dynamics; e.g. Fig.~\ref{f5a} shows that the radiated
energy is not given by the sum of integrated Larmor's rates. Another
attempt to check these interpretations is to correct the kinetic
energies in (\ref{eq:7}, \ref{eq:10}) by the ``cloud'' energies
\begin{eqnarray}
  \label{eq:200}
  \widetilde{K}=K-\frac{2e^2}{3c}
  \frac{\omega \dot{\omega}}{(1-\omega^2)^2}, ~~~
  \widetilde{K}'=K'-\frac{2e'^2}{3c}
  \frac{\omega' \dot{\omega}'}{(1-\omega'^2)^2},
\nonumber
\end{eqnarray}
and then to see whether leads to a better conservation law. We found
that the corrected kinetic energies do not generally lead to a better
conservation law than (\ref{eq:52}).

\comment{
\section{Clarification of (\ref{eq:52}, \ref{eq:522})}
\label{remarks}

Quantities in (\ref{kerman}, \ref{shah}) are calculated|in the
considered lab-frame|at the same times ($t_2$ and $t_1$,
respectively), but at different coordinates. One wonders whether this
is the only possibility, since events that are simultaneous in one
reference system will not be simultaneous in another
\cite{gamba}. (The most well-known example of such a quantity is the
length \cite{gamba}.)  Now, if it is expected that (say) a positive
amount of work comes from ${\rm P}'$ to ${\rm P}$, then one can try
the following definitions for the energy change of P:
  \begin{eqnarray}
    \label{eq:30}
&& \Delta E=E(t_2)-E(t_1),\\ 
&& \widetilde\Delta
  K'=K'(t_2-\delta'(t_2)\,)-K'(t_1-\delta'(t_1)\,),
  \end{eqnarray}
  and the kinetic energy change of ${\rm P}'$.  Their tentative
  rationale is that the delay in energy transfer is accounted for
  explicitly; cf. (\ref{eq:011}--\ref{eq:0111}). We studied this and
  similar quantities and found that effective conservation law
  $|\Delta E+\widetilde\Delta K'|\ll |\Delta E|,\, |\widetilde\Delta
  K'|$ either does not hold at all, or holds much worst than
  (\ref{eq:52}). We thereby stress once again that since the energy
  transfer is not defined a priori, the conservation relation
  (\ref{eq:52}) does proide an argument for choosing a better
  definition.

}


\clearpage

\begin{thebibliography}{99}

\bibitem{balian} R. Balian, {\it From Microphysics to Macrophysics},  
I (Springer, Berlin, 1992). 

\bibitem{lindblad}G. Lindblad, {\it Non-Equilibrium Entropy} {\it and
Irreversibility} (D. Reidel, Dordrecht, 1983).

\bibitem{mahler}
G. Mahler, {\it Quantum Thermodynamic Processes} (Pan Stanford,
Singapore, 2015).

\bibitem{lehrman}R.L. Lehrman, Phys. Teach. {\bf 11}, 15 (1973).

\bibitem{leff} H.S. Leff and A.J. Mallinckrodt, Am. J. Phys. {\bf 61},
  121 (1993).

\bibitem{mukamel}
M. Esposito, U. Harbola, and S. Mukamel, Rev. Mod. Phys. {\bf 81},
1665 (2009).


\bibitem{jarb}
C. Jarzynski, Eur. Phys. J. B {\bf 64}, 331 (2008).


\comment{Pellicane, G., Tsige, M., \& Aragie, B. (2015). Thermodynamics
  of a stochastic three level elevator model. The European Physical
  Journal B, 88(10), 1}

\bibitem{campisi}M. Campisi, P. Hanggi, and P. Talkner,
  Rev. Mod. Phys. {\bf 83}, 771 (2011).

\bibitem{skr}
    P. Skrzypczyk, A. J. Short, S. Popescu, Nat. Commun. {\bf 5},
    4185 (2014). 


\bibitem{armen}
A.E. Allahverdyan, Phys. Rev. E {\bf 90}, 032137 (2014).

\bibitem{gallego} 
R. Gallego, J. Eisert, and H. Wilming, {\it Defining
    work from operational principles}, arXiv:1504.05056.

\bibitem{aspects} 
P. Talkner and P. Hanggi, {\it Aspects of Work}, arXiv:1512.02516.

\bibitem{plastina}
F. Plastina {\it et al.},  Phys. Rev. Lett. {\bf
  113}, 260601 (2014).



\bibitem{3law}
A. E. Allahverdyan, K. V. Hovhannisyan, D. Janzing, and G. Mahler
Phys. Rev. E {\bf 84}, 041109 (2011).


\bibitem{rubi} J.M.G. Vilar, and J.M. Rubi, Phys.  Rev. Lett. {\bf
    100}, 020601 (2008); {\it ibid.} {\bf 101}, 098902 (2008); {\it
    ibid.} {\bf 101}, 098904 (2008).



\bibitem{peliti} L. Peliti, Phys. Rev. Lett. {\bf 101}, 098903 (2008).
  J. Stat. Mech.: Theory. Exp.  P05002 (2008).



\bibitem{silbey} E.N. Zimanyi and R. J. Silbey, J. Chem. Phys. {\bf
    130}, 171102 (2009).


\bibitem{rubirubi}J.M.G. Vilar, and J.M. Rubi,
J. Non-Equilib. Thermodyn. {\bf 36}, 123 (2011).


\bibitem{jar} J. Horowitz and C. Jarzynski,
Phys. Rev. Lett. {\bf 101}, 098901 (2008).


\bibitem{landau} L.D. Landau and E.M. Lifshitz, {\it The Classical
    Theory of Fields}, 4th ed. (Pergamon Press, Oxford, 1975).

\bibitem{kos}B. Kosyakov, {\it Introduction to the Classical Theory of
    Particles and Fields} (Springer, Berlin, 2007). 

\bibitem{yang}K.-H. Yang, Annals of Physics, {\bf 101}, 62 (1976).


\bibitem{kobe}D.H. Kobe, E.C.T. Wen and K.H. Yang, Phys. Rev. D {\bf
    26}, 1927 (1982).

D.H. Kobe and K.-H. Yang, J. Phys. A: Math. Gen. {\bf 13}, 3171 (1980).




\bibitem{zambrano} 
J. A. Sanchez-Monroy, J. Morales, and E. Zambrano, 
{\it Energy operator for non-relativistic and relativistic quantum
 mechanics revisited}, arxiv.org/1208.1425.

\bibitem{grif} D. J. Griffiths, Am. J. Phys. {\bf 80}, 7 (2012).

\bibitem{lorce_review} E. Leader and C. Lorc\'e, Phys. Rep.  {\bf
    541}, 163 (2014).



\bibitem{reiss_1}H. R. Reiss, {\it Limitations of gauge invariance},
  arXiv:1302.1212v1. 

\bibitem{reiss_2} H. R. Reiss, J. Mod. Optics {\bf 59}, 1371 (2012).


\bibitem{sipe} 
J. E. Sipe, Phys. Rev. A, {\bf 27}, 615 (1983)

\comment{

  New Hamiltonian for a charged particle in an applied electromagnetic
  field


We derive a new Hamiltonian, for a charged particle in a
time-dependent applied EMF which depends on the fields $E$ and $B$
directly rather than on the potentials $\phi$ and $A$. The Hamiltonian
is nonlocal in that it involves $E$ and $B$ at all points in space, in
contrast to the usual local Hamiltonian, which only involves $\phi$
and $A$ at the position of the charge. The new and usual Hamiltonians
are compared, and the canonical transformation which connects them is
presented.  We discuss the physical interpretation of the interaction
terms appearing in the canonical momentum, angular momentum, and the
Hamiltonian.  A relativistic generalization is given.}


\bibitem{wang_pra} W.-M. Sun, X.-S. Chen, X.-F. L\"u, and F. Wang,
  Phys. Rev. A, {\bf 82}, 012107 (2010).


\bibitem{feynman} R.P. Feynman, R.B. Leighton and M. Sands, {\it The
    Feynman Lectures on Physics} (Addison-Wesley, Reading, MA, 1964);
  chapter 27. 


\bibitem{slepian} J. Slepian, J. Appl. Phys. {\bf 13}, 512 (1942). 


C.S. Lai, Am. J. Phys. {\bf 49}, 841 (1981).


P. C. Peters,
Am. J. Phys. {\bf 50}, 1165 (1982).

R. H. Romer,
Am. J. Phys. {\bf 50}, 1166 (1982).

D. H. Kobe,
Am. J. Phys. {\bf 50}, 1162 (1982).


U. Backhaus and K. Schafer, Am. J. Phys. {\bf 54}, 279 (1986)


C.J. Carpenter, IEE Proc. {\bf 136}, 55 (1989).


\bibitem{jeffries}C. Jeffries, SIAM Rev. {\bf 34}, 386 (1992).






\bibitem{stewart} 
A. M. Stewart, Eur. J. Phys. {\bf 24}, 519 (2003). 


\comment{The vector decomposition theorem of Helmholtz leads to a form
  of the Coulomb gauge in which the potentials are expressed in a form
  that is totally instantaneous. The scalar potential is expressed in
  terms of the instantaneous charge density, the vector potential in
  terms of the instantaneous magnetic field.}


\bibitem{wundt}B.J. Wundt and U. D. Jentschura, Annals of Physics,
  {\bf 327} 1217 (2012).



\bibitem{heras_coulomb} J.A. Heras, Eur. J. Phys. {\bf 32}, 213
  (2011). 


J.A. Heras, Am. J. Phys. {\bf 75}, 176 (2007). 


\bibitem{heras_lorenz} J. A Heras and G. Fernandez-Anaya, {\it Can the
    Lorenz-gauge potentials be considered physical quantities?},
  arXiv:1012.1063v1.




\bibitem{dmitro} V. P. Dmitriyev,  Eur. J. Phys. {\bf 25}, L23 (2004).






\bibitem{puthoff} H.E. Puthoff, {\it Electromagnetic Potentials Basis
    for Energy Density and Power Flux}, arxiv.org/pdf/0904.1617.

\bibitem{bogo} N.N. Bogoliubov and D.V. Shirkov, {\it 
Introduction to theory of Quantized Fields} (J. Wiley, NY, 1980). 

\bibitem{fermi} E. Fermi, Rend. Lincei {\bf 9}, 881 (1929).

\bibitem{oost}A.B. van Oost, Eur. Phys. J. D, {\bf 8}, 9 (2000).









\bibitem{rouss_epl}G. Rousseaux, Europhys.
Lett. {\bf 71} 15 (2005).



\bibitem{lar}
J. Larsson, Am. J. Phys. {\bf 75}, 230 (2007). 



\bibitem{massphoton} L-C. Tu, J. Luo and G.T. Gillies,
    Rep. Prog. Phys. {\bf 68}, 77 (2005).



\bibitem{deffner} 
S. Deffner and A. Saxena, Phys. Rev. E {\bf 92}, 032137 (2015).



\bibitem{uribe} J.I. Jimenez-Aquino, F.J. Uribe and R.M. Velasco,
J. Phys. A {\bf 43}, 255001 (2010).



\bibitem{saha} A. Saha and A. M. Jayannavar, Phys. Rev. E {\bf 77},
  022105 (2008).


\bibitem{pradhan} P. Pradhan, Phys. Rev. E {\bf 81},
  021122 (2010).

\bibitem{yet_another} A. Saha, S. Lahiri, and A.M. Jayannavar Modern
  Physics Letters B {\bf 24}, 2899 (2010)


\bibitem{chub} A. Chubykalo, A. Espinoza, R. Tzonchev, Eur.
  Phys. J. D {\bf 31}, 113 (2004).  

\comment{ Experimental
    test of the compatibility of the definitions of the
    electromagnetic energy density and the Poynting vector

    It is shown that the generally accepted definition of the Poynting
    vector and the energy flux vector defined by means of the energy
    density of the electromagnetic field (Umov vector) lead to the
    prediction of the different results touching electromagnetic
    energy flux. The experiment shows that within the framework of the
    mentioned generally accepted definitions the Poynting vector
    adequately describes the electromagnetic energy flux unlike the
    Umov vector. Therefore one can conclude that a generally accepted
    definitions of the electromagnetic energy density and the Poynting
    vector, in general, are not always compatible.  }





\bibitem{synge}J.L. Synge, Proc. Roy. Soc. London A {\bf 177}, 118
  (1940).

\bibitem{stabler}R.C. Stabler, Phys. Lett. {\bf 8}, 185 (1964). 

\bibitem{cornish}F.H.J. Cornish, Proc. Phys. Soc. {\bf 86}, 427 (1965).

\bibitem{leiter} D. Leiter, J. Phys. A, {\bf 3}, 89 (1970).

\bibitem{beil} R.G. Beil, Phys. Rev. D, {\bf 12}, 2266 (1975)


\bibitem{franklin} J. Franklin and C. LaMont, Brazilian Journal of
  Physics {\bf 44}, 119 (2014).


\bibitem{kirpich} S.B. Kirpichev and P. A. Polyakov. Journal of
  Mathematical Sciences, {\bf 141}, 1051 (2007).


\bibitem{driver}R.D. Driver, Ann. Phys. {\bf 21}, 122 (1963).

\bibitem{driver_existence} R.D. Driver and M.J. Norris, Ann. Phys.
  {\bf 42}, 347 (1967).

  \comment{ Note on uniqueness for a one-dimensional two-body problem
    of classical electrodynamics.

    A previous paper [Ann. Phys.21, 122 (1963)] gave a mathematical
    analysis of a two-body problem of classical electrodynamics
    incorporating retarded interactions (for the case of two charged
    particles remaining on the x axis). It was shown that the future
    trajectories would be uniquely determined, and well behaved, if
    past histories of the particles were given satisfying certin
    conditions, including a Lipschitz condition on the
    velocities. However, this Lipschitz condition has since become
    suspect as possibly unreasonable. The present note shows that it
    was, in fact, also superfluous.
}

\bibitem{driver_book}R. D. Driver, {\it Ordinary and Delay
    Differential Equations} (Springer-Verlag, New York, 1977).

\bibitem{book_numerics}A. Bellen and M. Zennaro, {\it Numerical
    methods for delay differential equations} (Oxford University
  Press, Oxford, 2003).

\bibitem{raju}C.K. Raju, {\it Time: Towards a Consistent Theory}
  (Kluwer Academic, Dordrecht, 1994).


\bibitem{baylis}J. Huschilt, W. E. Baylis, D. Leiter, and G. Szamosi,
  Phys. Rev. D {\bf 7}, 2844 (1973).


\bibitem{kasher} J. C. Kasher, Phys. Rev. D {\bf 12}, 1729 (1975).

\comment{
  On the formulation of initial-value problems for systems consisting
  of relativistic particles.

  We discuss questions related to the well-posedness of problems on
  the motion of relativistic many-body systems. For one-dimensional
  relativistic motion of N similar charges, we prove that an ordinary
  Cauchy problem usual in Newton mechanics can be stated; this is done
  in the framework of microscopic Maxwell-Lorentz electrodynamics
  (including a model with self-action) or Wheeler-Feynman theory.

}

\bibitem{driver_back} R.D. Driver, Phys. Rev. {\bf 178}, 2051 (1969).

\bibitem{travis}S.P. Travis, Phys. Rev. D {\bf 11}, 292 (1975).


\bibitem{hsing} D.-P. Hsing, Phys. Rev. D {\bf 16}, 974 (1977).

\bibitem{zhdanov} V.I. Zhdanov, J. Phys. A: Math. Gen. {\bf 24}, 5011
  (1991).

\bibitem{aichelburg}P.C. Aichelburg and H. Grosse, 
Phys. Rev. D {\bf 16}, 1900 (1977).




\comment{ 

\bibitem{persides}
S. Persides and J. Pascalis, Annals of Physics, {\bf 87} 161 (1974).

161-175. The electromagnetic two-body problem in special relativity

  A system of two point charged particles is considered. Each particle
  moves in the electromagnetic field created by the other particle
  according to Maxwell's equations. A scheme of successive
  approximations is developed to study the field and the motion of the
  charges. The field (potentials and intensities) are exapanded in
  powers of $c^{-1}$ using a retarded time coordinate. The variables
  of the motion (position vectors, velocities, etc) are expanded in
  powers of $c^{-1}$ with coefficients depending on t only. The field
  is evaluated in the first three approximations. The equations of
  motion are derived in the same approximations and the corresponding
  conserved quantities are explicitly given. Thus, the usual
  assumption of an action-at-a-distance principle is avoided and the
  original nonlinear integrodifferential equations are reduced to a
  sequence of linear equations.  

}

\bibitem{bel}L. Bel and J. Martin, Phys. Rev. D {\bf 8}, 4347 (1973).





\comment{
\bibitem{bizarro} 
J. P. S. Bizarro, Phys. Rev. E, {\bf 78}, 021137 (2008).

Entropy production in irreversible processes with friction

Established expressions for entropy production in irreversible
processes are generalized to include friction explicitly, as a source
of irreversibility in the interaction between a system and its
surroundings. The net amount of heat delivered to the system does not
come now only from the reservoir, but may have an additional component
coming from the work done against friction forces and dissipated as
heat. To avoid ambiguities in interpreting the different contributions
to entropy increase, the latter is also written in terms of the heat
directly exchanged between the system and surroundings and of the
fraction of frictional work that is lost in the system. }


\bibitem{pugh}
E. M. Pugh and G. E. Pugh, Am. J. Phys. {\bf 35}, 153 (1967).






\bibitem{fock}
V. Fock and B. Podolsky. Sov. Phys {\bf 1}, 801 (1932).


\bibitem{gritsunov}A.V. Gritsunov, Radioelectronics and
  Communications Systems, {\bf 52}, 649 (2009). 


\bibitem{babi} L. Allen, M.J. Padgett, and M. Babiker, Progress In
  Optics, {\bf XXXIX}, 39, 291 (1999).


\bibitem{gamba}A. Gamba, Am. J. of Phys. {\bf 35}, 83 (1967).

\bibitem{str} V.N. Strel'tsov, Foundations of Physics, {\bf 6}, 293
  (1976).

\bibitem{trautman} A. Trautman, Nature, {\bf 242}, 7 (1973).


\bibitem{sciama}D.W. Sciama, Mathematical Proceedings of the Cambridge
  Philosophical Society, {\bf 54}, No. 01 (Cambridge University Press,
  1958).

  \comment{ On a non-symmetric theory of the pure gravitational field

    It is suggested, on heuristic grounds, that the energy-momentum
    tensor of a material field with non-zero spin and non-zero
    rest-mass should be non-symmetric. The usual relationship between
    energy-momentum tensor and gravitational potential then implies
    that the latter should also be a non-symmetric tensor. This
    suggestion has nothing to do with unified field theory; it is
    concerned with the pure gravitational field.  A theory of
    gravitation based on a non-symmetric potential is developed. Field
    equations are derived, and a study is made of Rosenfeld
    identities, Bianchi identities, angular momentum and the equations
    of motion of test particles. These latter equations represent the
    geodesics of a Riemannian space whose contravariant metric tensor
    is gij, in agreement with a result of Lichnerowicz on the
    bicharacteristics of the Einstein Schrï¿½dinger field equations.}

\bibitem{bregman} K. Hayashi and A. Bregman, Annals of Physics, {\bf
    75}, 562 (1973).


\bibitem{thermo_rel} Relativistic statistical thermodynamics is a vast and
  controversial subject with many contributions over the last 110
  years. We cite only three recent reviews:

  M. Requardt, {\it Thermodynamics meets Special Relativity - or what
    is real in Physics?} arXiv preprint arXiv:0801.2639.

J. Dunkel and P. Hanggi, Phys. Rep. {\bf 471}, 1 (2009). 


M. Przanowski and J. Tosiek, Physica Scripta, {\bf 84},
    055008 (2011).



  \bibitem{gabovich} A.M. Gabovich and N. A. Gabovich,
    Eur. J. Phys. {\bf 28}, 649 (2007).


\bibitem{glass} E. N. Glass, J. Huschilt, and G. Szamosi,
  Am. J. Phys. {\bf 52}, 445 (1984).

\bibitem{baylis_retarded}
J. Huschilt, W. E. Baylis, Physical Review D {\bf 13}, 3256 (1976).


\bibitem{silenko}
A. J. Silenko, Physical Review A {\bf 91}, 012111 (2015). 
\comment{
 Energy expectation values of a particle in nonstationary fields

We show that the origin of the nonequivalence of Hamiltonians in
different representations is a change of the form of the
time-derivative operator at a time-dependent unitary
transformation. This nonequivalence does not lead to an ambiguity of
the energy expectation values of a particle in nonstationary fields
but assigns the basic representation. It has been explicitly or
implicitly supposed in previous investigations that this
representation is the Dirac one. We prove the alternative assertion
about the basic role of the Foldy-Wouthuysen representation. We also
derive the general equation for the energy expectation values in the
Dirac representation. As an example, we consider a spin-1/2 particle
with anomalous magnetic and electric dipole moments in strong
time-dependent electromagnetic fields. We apply the obtained results
to a spin-1/2 particle in a plane monochromatic electromagnetic wave
and give an example of the exact Foldy-Wouthuysen transformation in
the nonstationary case. 

}


\bibitem{russo}G. Rousseaux, Ann. Fond. Louis de Broglie {\bf 28},
  261 (2005).



\bibitem{blondel}
G. Giuliani, Eur. J. Phys. {\bf 31}, 871 (2010).

\comment{
Vector potential, electromagnetic induction and "physical meaning"


A forgotten experiment by Andr\'e Blondel (1914) proves, as held on
the basis of theoretical arguments in a previous paper, that the time
variation of the magnetic flux is not the cause of the induced emf:
the physical agent is instead the vector potential through the term
$-\partial A /\partial t$ (when the induced circuit is at rest). The
``good electromagnetic potentials'' are determined by the Lorenz
condition and retarded formulas. Other pairs of potentials derived
through appropriate gauge functions are only mathematical devices for
calculating the fields: they are not physically related to the
sources. The physical meaning of a theoretical term relies, primarily,
on theoretical grounds: a theoretical term has physical meaning if it
cannot be withdrawn without reducing the predictive power of a theory
or, in a weaker sense, if it cannot be withdrawn without reducing the
descriptive proficiency of a theory.}

\bibitem{bobrov} V.B. Bobrov, S.A. Trigger, G.J.F. van Heijst,
  P.P.J.M. Schram, {\it Aharonov-Bohm effect, electrodynamics
    postulates, and Lorentz condition}, arXiv:1306.6736.

\end{thebibliography}
\end{document}